\documentclass[12pt]{JHEP3}% 10pt is ignored!

\usepackage{epsfig,multicol,amsmath}
\newcommand\fverb{\setbox\pippobox=\hbox\bgroup\verb}
\newcommand\fverbdo{\egroup\medskip\noindent%
\fbox{\unhbox\pippobox}\ }			
\newcommand\fverbit{\egroup\item[\fbox{\unhbox\pippobox}]}
\newbox\pippobox
\def\d2bar{$\overline{\mbox D2}$}
\title{Living Near de Sitter Bubble Walls}
\author{Jin-Ho Cho$^{1,2}$ and Soonkeon Nam$^{1}$\\
$^{1}$Department of Physics \& Research Institute for Basic Sciences,\\ Kyung Hee University, Seoul 130-701, Korea\\
$^{2}$Center for Quantum Space Time, Sogang University, Seoul 121-742, Korea\\\\
E-mail: \email{cho.jinho@gmail.com}, \email{nam@khu.ac.kr}}

\preprint{\hepth{0607098}} % OR: \preprint{Aaaa/Mm/Yy\\Aaa-aa/Nnnnnn}
\abstract{We study various bubble solutions in string/M theories obtained by double Wick rotations of  (non-)extremal brane configurations. Typically, the geometry interpolates de Sitter space-time $\times$ non-compact extra-dimensional space in the near-bubble wall region and the asymptotic flat Minkowski space-time. These bubble solutions provide nice background geometries reconciling string/M theories with de Sitter space-time. For the application of these solutions to cosmology, we consider multi-bubble solutions and find landscapes of varying cosmological constant. Double Wick rotation in string/M theories, used in this paper, introduces imaginary higher-form fields. Rather than regard these fields as {\em classical} pathologies, we interpret them as {\em semi-classical} decay processes of de Sitter vacuum via the production of spherical branes. We speculate on the possibility of solving the cosmological constant problem making use of the condensation of the spherical membranes.}

\keywords{bubble, de Sitter space-time, string/M-theories, creation of spherical branes,  cosmological constant problem}

\begin{document} 
\section{Introduction}
\subsection{De Sitter Space-time and String Theory}
Nearing the end of last century, astronomers had an astonishing discovery, the universe is expanding with acceleration \cite{Perlmutter:1998np}\cite{Riess:1998cb}. This discovery called for many new ideas, since the conventional gravity is an attractive force and cannot accelerate the expansion of the Universe. 

Positive cosmological constant \cite{Carroll:2000fy}\cite{Padmanabhan:2002ji}\cite{Weinberg:1988cp}\cite{Polchinski:2006gy} is the oldest and simplest solution, but for string theorists, poses a daunting problem. De Sitter space-time which is the cosmological solution with positive cosmological constant $\Lambda$, does not easily allow string theory in it \cite{Balasubramanian:2004wx}.
The problem is as follows. There is a deep rooted problem of formulating quantum field theory in de Sitter space-time, let alone string theory. The basic reason is that whereas de Sitter space-time only allow finite degrees of freedom, quantum field theory (or string theory) has infinite degrees of freedom. The finiteness of degrees of freedom in de Sitter space-time is related to the fact that it has finite entropy and thus the degrees of freedom in it can only be finite \cite{Banks:2000fe}\cite{Fischler:2001yj}\cite{Witten:2001kn}\cite{Bousso:2002fq}\cite{Goheer:2002vf}. There is no unitary way to act de Sitter group on the finite degrees of freedom.

To make matters worse, there is a `no-go theorem' in string theory (supergravity) which rules out de Sitter space-time as a result of non-singular warped compactification \cite{Maldacena:2000mw}. The basic assumptions for this theorem are i) no higher curvature correction in the gravity action, ii) the non-positive potentials for the scalars, iii) massless fields with positive kinetic terms, and iv) finite effective Newton's constant in lower dimensions.

In this paper, we study various bubble solutions in string/M theories and their applications to cosmological problems. The bubble solutions have a region, the near-bubble wall region, which is de Sitter, but asymptotically are flat Minkowski. Therefore, they have plenty of rooms to allow infinite degrees of freedom of string theory. Moreover, we will argue that most matters are accumulated near the bubble wall as the bubble expands. Therefore most of the galaxies are formed near the bubble wall and it is natural that intelligent observers see de Sitter space-time around them.

These solutions are obtained by performing the double Wick rotation (DWR) on well-known D/M-brane solutions and we will call them D/M-bubbles. First, we find that near the bubble wall, the geometry becomes `de Sitter $\times$ non-compact internal space'. Second, we find that the case of extremal D3-bubble corresponds to Hull's Euclidean brane (called E4-brane) \cite{Hull:1998vg}. However, other cases we have considered, such as M2- or M5-bubbles are new and cannot be obtained by Hull's timelike T-duality of type II theories because there is no notion of T-duality in M theory. 

Extremal D$3$-bubbles (including M$2$- and M5-bubbles) preserve full 32 supersymmetries near the bubble wall and at the asymptotic infinity. In order to make sense of supersymmetry in this background, Ramond-Ramond (RR) should be imaginary, which is also a consequence of DWR. Imaginary valued fields, at least for free cases, can be regarded as fields with the wrong sign for the kinetic term. At first glance, this might sound pathological. However, there is a long history of fields with the wrong kinetic term \cite{Hoyl}\cite{Gibbons:2003yj}, and recently such fields have been considered as the candidate for the dark energy \cite{Caldwell:1999ew} which are the source for the acceleration of the expansion of the Universe. (However, certain problems about the kinetic terms with wrong sign have been pointed out in view of phenomenology \cite{Carroll:2003st}.)

In this paper, we {\em assume the imaginary valued fields at the semi-classical level} and pursue the foregoing argument about de Sitter space-time in 10- or 11-dimensional supergravity. Although this scheme is quite against the standard folklore, one should have checked at least whether it leads to any inconsistency or unphysical results. One basic `pro' for it is that the string theory does not exclude the possibility of the imaginary valued fields from the beginning. Let us consider a string field state in the string Fock space basis as
\begin{equation}\label{e1}
|\Psi>=\int d^{D}p \left( \phi(p)|p>+t_{\mu\nu}(p)\alpha^{\mu}_{-1}\tilde{\alpha}^{\nu}_{-1}|p>+\cdots\right),
\end{equation}  
where $\alpha^{\mu}_{-n}=(\alpha^{\mu}_{n})^{\dag}$, and so on, to ensure Hermiticity of the string coordinate field $X^{\mu}=X^{\mu\dag}$. When we discuss excitations over time-independent vacua, we restrict the coefficients, $\phi(p),\,t_{ \mu\nu}(p),\cdots$, to be real. However, in priniciple, the coefficients can be complex without causing any logical inconsistency. There is no {\it a priori} reason to restrict the string field state to the one in real Fock space.   

On the other hand, one might easily raise a naive `con' too, against the imaginary valued fields. The situation gets more complicated if we couple the imaginary field with real valued charges. The coupling gives rise to an imaginary term in the Lagrangian in constrast with the real (though wrong signed) kinetic term. Our proposal is that we should interpret such a term in a semi-classical way as a signature of the instability of the vacuum. This is in the spirit of Schwinger's original work \cite{Schwinger:1951nm} on the pair creation in a strong electric field. Here, the imaginary value of the effective potential has a definite physical interpretation. It induces the creation of spherical branes much like Schwinger's process of pair creation.

This paper is one attempt to incorporate de Sitter spacetime into string/M theories and we will see the ghost fields appearing via DWR actually do not   
any serious harm when the story involves an unstable vacuum like de Sitter\footnote{De Sitter space-time is classically stable in the sense that the cosmological constant persists its value with time. However, there could be some quantum instability like particle creation \cite{Mottola:1984ar} or decaying to other de Sitter \cite{Ford:1984hs}\cite{Mottola:1985qt}\cite{Antoniadis:1985pj}\cite{Barrow:1986yf}.}. It rather gives us a new insight about de Sitter space-time in the language of brane physics.  Furthermore, the bubble solutions have a surprising cosmological application. 

The empirical evidences (the acceleration of the present universe and the dominant contribution from the matter component with the equation of state parameter $w\sim -1$ \cite{Riess:1998cb}) for the positive cosmological constant impose on us the problem of explaining why the cosmological constant is so small but is yet non-vanishing. This `cosmological constant problem' is notoriously difficult to solve. In this paper, we address this problem again in the context of string/M theories and find a surprisingly simple picture of the solution, which is generically quantum mechanical and also generically string theoretic. This is achieved by utilizing the created spherical branes as a mechanism of lowering the cosmological constant of de Sitter space-time.

\subsection{Detouring the No-go theorem in Supergravity}

There are several ways to detour the no-go theorem  \cite{Maldacena:2000mw}. For example, one can use some negative tension objects like O$3$-planes \cite{Giddings:2001yu} or conceive of the compactification on some `non-compact' internal space like hyperbolic space \cite{Townsend:2003fx}: This specific compactification can happen in the geometry around `S-branes' \cite{Ohta:2003pu}.\footnote{It was also stressed in Refs. \cite{Chen:2003dc} \cite{Ohta:2003ie} that the accelerating feature of the solutions is concerned rather with the time-dependent compactification than with the hyperboloidal structure of the internal manifold. See also Refs. \cite{Neupane:2005ms} \cite{Neupane:2005nb} for this.} In the same vein, quantum corrections and extended objects in the warped geometry provide a controllable way to get de Sitter space-time while stabilizing all the moduli \cite{Kachru:2003aw}.  Although all these schemes were quite successful in getting to de Sitter space-time in the framework of string/M theories, we are still lost ironically in a vast landscape of de Sitter vacua \cite{Susskind:2003kw}.

In this paper, we suggest another way of detouring the no-go theorem; i.e., introducing imaginary fields, or equivalently real fields with wrong kinetic terms. One virtue of introducing the imaginary fields is that de Sitter space-time can be obtained in a very simple setup as an exact solution of Einstein equation. For example, let us take $4$-dimensional Einstein-Maxwell system. The equations of motion (written in the orthonormal frame) taking the form,
\begin{equation}\label{e2}
R_{ab}= \frac{1}{2}F_{ac}F_{b}{}^{c}- \frac{1}{8}\eta_{ab}F_{cd}F^{cd},\qquad d*F^{(2)}=0
\end{equation}  
have the well-known  Freund-Rubin type solution,
\begin{equation}\label{e3}
ds^{2}=ds^{2}_{ { AdS}}+ ds^{2}_{ {sphere}}, \qquad F^{(2)}=m e^{2}\wedge e^{3},
\end{equation}  
where $e^{2}\wedge e^{3}$ is the volume of the unit sphere. From the observation that the nontrivial components of Ricci tensor are $R_{00}=-R_{11}=R_{22}=R_{33}=m^{2}/4$, we are tempted to think of an imaginary valued field strength ($m^{2}<0$) so that the above solution get converted to the type
\begin{equation}\label{e4}
ds^{2}=ds^{2}_{ {hyperbolic}}+ ds^{2}_{ {dS}},
\end{equation}   
that is, the geometry factorized into two-dimensional hyperbolic space and two-dimensional de Sitter space-time.

%\subsection{The Double Wick Rotation as a Symmetry}

A systematic way of making imaginary fields is the procedure of DWR. In contrast to the familiar single Wick rotation, (that is the analytic continuation to Euclidean space), DWR is a true symmetry at the action level. See Appendix A for a detailed proof. A typical form of DWR is

\begin{equation}\label{e5}
t = -i\xi, \quad x = i\psi,
\end{equation}    
which corresponds to the simultaneous Wick rotation of the temporal coordinate $t$ and one of the spatial coordinates, $x$.
Actually the Jacobian factor involved in this DWR is trivial and we need not define a new  (DWR) version of the Lagrangian. Recall that in the conventional single Wick rotation procedure, we 
define Euclidean Lagrangian as
\begin{equation}\label{e6}
\mathcal{L}_{E}\left(q,\,\frac{dq}{d t_{E}}\right)\equiv-\mathcal{L}\left(q,\,i\frac{dq}{d t_{E}}\right), 
\end{equation}   
that is minus of Minkowskian Lagrangian with the replacement $t=-i t_{E}$.

Once we are given a set of solutions of the equations of motion, this discrete symmetry of DWR can be used to generate another set of solutions. This machinery works because the supergravity action is intact under DWR, therefore,  the newly generated solutions solve the original equations of motion with appropriate boundary conditions.

\subsection{Geometry near the Wall of Witten's Bubble}

One typical example of thus made solution is the original Witten's bubble solution \cite{Witten:1981gj}, obtained by double Wick rotating Schwarzschild black hole solution. Various aspects of bubble solutions have been considered in Refs. \cite{Aharony:2002cx}\cite{Balasubramanian:2005bg}\cite{Astefanesei:2005eq}\cite{LopezCarballo:2005vp}. With string/M theories in mind, let us consider Schwarzschild black hole in general $D$-dimensions;
\begin{equation}\label{e7}
ds^{2}=- \left(1- \left( \frac{r_{0}}{r}\right)^{D-3}  \right)dt^{2}+\left(1- \left( \frac{r_{0}}{r}\right)^{D-3}  \right)^{-1}dr^{2}+r^{2} \left( d\theta^{2}+\sin^{2}{\theta}d\Omega^{2}_{D-3}\right). 
\end{equation}
Taking the following double Wick rotation,
\begin{equation}\label{dwr}
t=-i\xi, \qquad \theta= \frac{\pi}{2}+i\psi,
\end{equation}  
we obtain a bubble solution
\begin{equation}\label{wittenbubble}
ds^{2}= \left(1- \left( \frac{r_{0}}{r}\right)^{D-3}  \right)d\xi^{2}+\left(1- \left( \frac{r_{0}}{r}\right)^{D-3}  \right)^{-1}dr^{2}+r^{2} \left( -d\psi^{2}+\cosh^{2}{\psi}\,d\Omega^{2}_{D-3}\right). 
\end{equation}

The radial coordinate should be restricted as $r\ge r_{0}$, to prevent three temporal coordinates from appearing. Regularity at $r=r_{0}$ requires that the coordinate $\xi$ be periodic;
\begin{equation}\label{e8}
\xi\sim \xi+ \frac{4\pi r_{0}}{D-3}.
\end{equation}
Asymptotically, the solution describes the standard Kaluza-Klein compactification. This is more transparent in Rindler  coordinates,
\begin{equation}\label{rindler}
\tau=r\sinh{\psi},\qquad \rho=r\cosh{\psi},\qquad (\tau^{2}<\rho^{2})
\end{equation}  
in which the asymptotic geometry becomes a circle times $(D-1)$-dimensional flat Minkowski space-time;
\begin{eqnarray}\label{e9}
ds^{2}&\simeq&d\xi^{2}+dr^{2}+r^{2} \left( -d\psi^{2}+\cosh^{2}{\psi}\,d\Omega^{2}_{D-3}\right) 
 \nonumber\\
&=&d\xi^{2}-d\tau^{2}+d\rho^{2}+\rho^{2}\,d\Omega^{2}_{D-3}\,.
\end{eqnarray}    
The internal circle shrinks to zero size at $r=r_{0}$ and the geometry is not extended to the region $r<r_{0}$, thus it is called a bubble of nothing. The whole nowhere-singular solution describes a bubble of nothing with its wall accelerating in time ($\rho^{2}=\tau^{2}+r^{2}_{0}$). In this sense, the solution is also called as Kaluza-Klein bubble. Fig.~1 shows the snapshot of a bubble. In the asymptotic region, the geometry represents a flat Minkowski space-time Kaluza-Klein compactified on a circle of a period $\Delta\xi$.

\FIGURE{
\epsfig{file=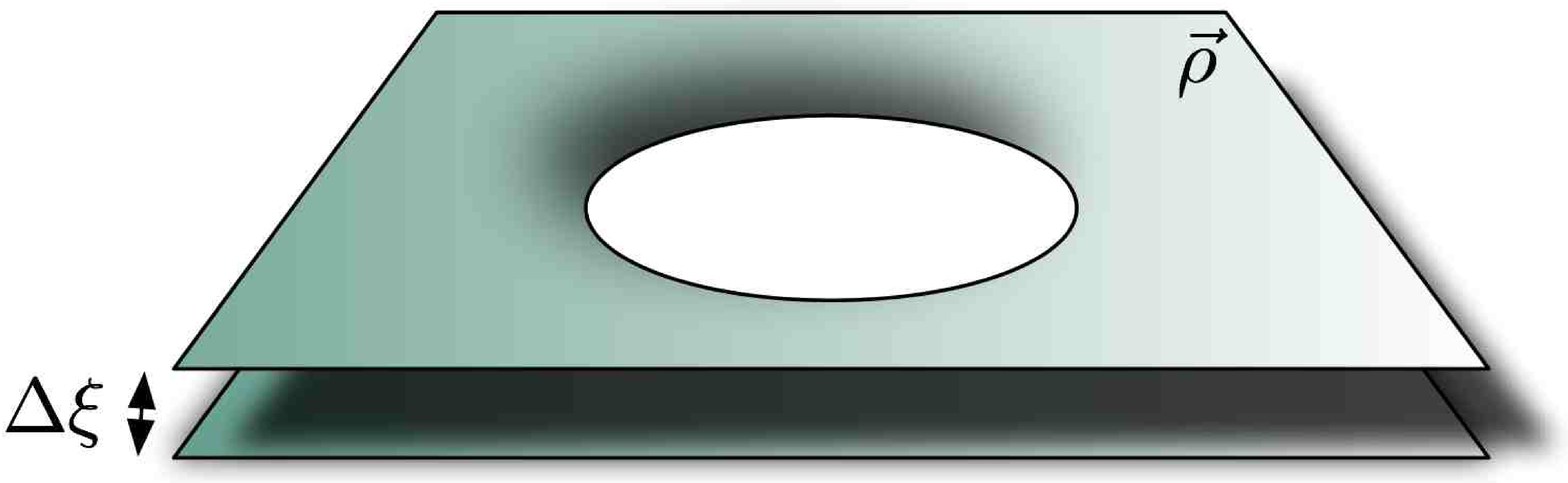,width=10cm} 
        \caption{A Kaluza-Klein Bubble}
	\label{figure1}}

%\begin{figure}[htbp]
%\begin{center}
%\includegraphics[width=11cm]{penrose.jpg}
%\caption{A Kaluza-Klein Bubble}
%\label{figure1}
%\end{center}
%\end{figure}
Since the asymptotic isometry of the bubble solution is $U(1)\times SO(D-2,\,1)$,
one is tempted to think of the `near-bubble' geometry, where $(D-2)$-dimensional de Sitter space-time (with the isometry $SO(D-2,\,1)$) could be relevant. This looks plausible and the approximate geometry in the near-bubble region describes a disk and a $(D-2)$-dimensional de Sitter space-time;
\begin{equation}\label{e10}
ds^{2}\simeq \frac{1}{r^{D-3}_{0} } \left(u^{2}d\xi^{2}+ \frac{4r^{2}_{0}}{\left(D-3 \right)^{2} } du^{2} \right)+r^{2}_{0} \left(-d\psi^{2}+\cosh^{2}{\psi}\,d\Omega^{2}_{D-3} \right).  
\end{equation}  
Here, we introduced the near-bubble coordinate $u^{2}=r^{D-3}-r^{D-3}_{0}$ and neglected the terms of order $\mathcal{O}(u^{2}/r^{D-3}_{0}) $. However, we should not take this `factorization' seriously because this approximate geometry does not solve the equation of motion exactly\footnote{If there is some mechanism that makes the observer remain at fixed point of the radial coordinate $u$, then he will find himself in de Sitter space-time. This can be achieved by giving some angular momentum to the observer. In fact, this possibility was studied in Ref. \cite{Brill:1989rp} by considering the geodesic motions in a Kaluza-Klein bubble background.}.

In this paper, we apply DWR to various D-brane and M-brane solutions well-known in string/M theories and get their bubble cousins. The non-extremal bubbles exhibit similar features to the Witten's bubble's; the solutions are restricted to the outside of some fixed radial coordinate $r_{0}> 0$, the asymptotic geometry describes Minkowski space-time compactified on a circle, and the bubble wall is accelerating in the asymptotic coordinates. 

The extremal bubbles inherit most good features from their D-brane and M-brane counterparts. The geometry near the bubble wall, sharply factorized as de Sitter space-time and the hyperbolic space, is not just an approximate solution but an exact solution. The wall of the extremal bubbles follows a null line, nevertheless the near bubble solution preserves the full $32$ supersymmetries. 

However, those solutions are plagued by various imaginary valued fields therefore look `pathological'. This is the consequence of the DWR from the real valued fields of AdS$_{p}\times$ S$^{q}$ backgrounds. The option one can choose at this point is either to abandon those solutions or to interpret their physical meaning. The strategy we will follow in this paper is the latter. We interpret the DWR procedure as a physical one; the de Sitter part including imaginary gauge fields as the instanton tunneling from S$^{4}$ to dS$_{3+1}$ \`{a} la Coleman-de Luccia \cite{Coleman:1980aw} or Hawking-Turok \cite{Hawking:1998bn} in the spirit of Hartle-Hawking no-boundary proposal \cite{HartleHawking}. We would like to emphasize here that our solutions need not assume any false vacua as was done in Coleman-de Luccia instanton and  are also in contrast with Hawking-Turok instanton in that they are geodesically complete without any singularity.     

This semi-classical interpretation of the bubble solutions is not that unreasonable. Recall the motion of a particle in one dimension under a potential $V(x)$. For a given energy $E$ of the particle, its motion under the potential barrier (the region where $E<V(x)$) is forbidden in the classical sense because there, its momentum $p\sim\sqrt{E-V(x)}$ becomes purely imaginary valued. We have no classical way to measure this imaginary momentum. However, the particle can actually be found under the potential barrier with a finite probability unless the barrier is infinitely high. At the quantum level, the measurement of the imaginary momentum is performed by measuring the probability of the particle to be under the barrier because WKB wave function of the particle is of the form $\psi(x)\sim e^{i\int^{x}\,p(x')\,dx'}$ with $p(x')$ imaginary valued \cite{Sakurai}. The same physics of the tunneling process can be read via Euclidean method with purely real Euclidean momentum achieved by Wick rotation \cite{Coleman}. 
Though we obtained the bubble solutions solving the {\it classical} equations of motion, the way they obtained is semi-classical (double Wick rotation). As a result, we are cluttered by the imaginary values. The way we understand their physical meaning should also be semi-classical. 

\section{D-bubbles and M-bubbles}

\subsection{D/M-brane Configurations and their Bubble Cousins}
Most well-known D/M-brane configurations have their bubble cousins. In this section, we will study in detail the bubble solution obtained from a D$3$-brane solution. We will call it a D3-bubble. Bubble solutions can also be obtained from other well-known brane configurations, such as D1-D5. We can also study bubbles from M-branes, thus called M-bubbles. Details of these other bubble solutions can be found in the appedix B.

Let us start with a non-extremal D3-brane solution;
\begin{eqnarray}\label{e11}
ds^{2}
&=&H_{3}^{ -\frac{1}{2}}(r) \left(-f_{3}(r)dt^{2}+d\vec{x}^{2}  \right)+H_{3}^{ \frac{1}{2}}(r) \left( f_{3}^{-1}(r)dr^{2}+r^{2}d\Omega^{2}_{5}\right), \nonumber\\ 
H_{3}(r)&=& 1+ \frac{\mu_{3}^{4}\sinh^{2}\alpha_{3}}{r^{4}}, \qquad f_{3}(r)=1- \frac{\mu_{3}^{4}}{r^{4}} 
\end{eqnarray}
with the electric and the magnetic RR field strength as
\begin{eqnarray}\label{e12}
\mathcal{F}_{(5)}&=&-*dH_{m}+d \left(H^{-1}_{e}-1 \right)\wedge dt\wedge dx_{1}\wedge dx_{2}\wedge dx_{3} \nonumber\\
&=&4\mu^{4}_{3}\,\sinh\alpha_{3}\cosh\alpha_{3}\,\left(d\Omega_{5}+r^{-5}H^{-2}_{3}dt\wedge dx_{1}\wedge dx_{2}\wedge dx_{3}\wedge dr \right) ,\nonumber\\
H_{m}&=&1+ \frac{\mu^{4}_{3}\sinh\alpha_{3}\cosh\alpha_{3}}{r^{4}},\quad H^{-1}_{e}-1=- \frac{\mu^{4}_{3}\sinh{\alpha_{3}}\cosh{\alpha_{3}}}{r^{4}+\mu^{4}_{3}\sinh^{2}{\alpha_{3}}}.
\end{eqnarray}   
Here, Hodge star $*$ is with respect to the $6$-dimensional {\it flat} space  transverse to the brane world-volume.  We followed the notations for the harmonic functions from Ref. \cite{Cvetic:1996gq}. Throughout this paper, we assume the parameters concerning RR charges are very large in the string unit so as to make the supergravity solutions trustable. For example, in Eq. (\ref{e12}), $\mu^4_3\sinh\alpha_3\cosh\alpha_3\gg\alpha'^2$.

The bubble geometry is obtained \`{a} la Witten (\ref{dwr}) \cite{Witten:1981gj}.
Double Wick rotations along the temporal direction and one of the spherical direction  lead us to 
\begin{equation}\label{d3bubble1}
ds^{2}=H_{3}^{ -\frac{1}{2}}(r) \left(f_{3}(r)d\xi^{2}+d\vec{x}^{2}  \right)+H_{3}^{ \frac{1}{2}}(r) \left( f_{3}^{-1}(r)dr^{2}+r^{2} \left(\cosh^{2}\psi\, d\Omega^{2}_{4} -d\psi^{2}\right)\right).
\end{equation}
The fact that RR field strength is involved in the solution is basically in contrast with Witten's bubble (\ref{wittenbubble}). It results in the `unwanted' imaginary value of the field;
\begin{equation}\label{imaginaryf5}
\mathcal{F}_{(5)}=4i\mu^{4}_{3}\,\sinh\alpha_{3}\cosh\alpha_{3}\left(\cosh^{4}\psi\, d\Omega_{4} \wedge d\psi -r^{-5}H^{-2}_{3}d\xi\wedge dx_{1}\wedge dx_{2}\wedge dx_{3}\wedge dr\right).
\end{equation}    
However, as was mentioned earlier, we will interpret this imaginary value of the solution as some quantum process like instanton. 

\subsection{Near-Bubble Geometries and de Sitter space-time}
Here, we will see that near bubble geometries contain de Sitter space-time.
In the near-bubble coordinate $(u^{2}=r^{4}-\mu_{3}^{4},\quad u^{2}\ll \mu_{3}^{4})$, the harmonic functions are well approximated by
\begin{eqnarray}\label{e13}
f_{3}(u)&=& \frac{u^{2}}{u^{2}+\mu_{3}^{4}}\simeq \frac{u^{2}}{\mu_{3}^{4}}
\nonumber\\
H_{3}(u)&=&\frac{u^{2}+\mu_{3}^{4}\cosh^{2}\alpha_{3}}{u^{2}+\mu_{3}^{4}}\simeq\cosh^{2}\alpha_{3}- \sinh^{2}{\alpha_{3}}\frac{u^{2}}{\mu^{4}_{3}} \\
H_{m}(u)&=&\frac{u^{2}+\mu_{3}^{4} \left(1+\sinh\alpha_{3}\cosh\alpha_{3} \right)}{u^{2}+\mu_{3}^{4}}\simeq 1+\sinh\alpha_{3}\cosh\alpha_{3}-\sinh\alpha_{3}\cosh\alpha_{3}\frac{u^{2}}{\mu^{4}_{3}}. \nonumber
\end{eqnarray}
As a consequence, the metric takes the form;
\begin{eqnarray}\label{e14}
ds^{2}&\simeq& \frac{\cosh\alpha_{3}}{4\mu_{3}^{2}} \left(du^{2}+ \frac{4u^{2}}{\mu_{3}^{2}\cosh^{2}\alpha_{3}}d\xi^{2} \right)+\cosh^{ -1}\alpha_{3} \sum^{3}_{x=1}dx_{i}^{2}\nonumber\\
&&\qquad\qquad+\mu_{3}^{2}\,\cosh\alpha_{3}\, \left(-d\psi^{2}+\cosh^{2}\psi\,d\Omega_{4}^{2} \right),   \nonumber\\
\mathcal{F}_{(5)}&\simeq& 4i\mu^{4}_{3}\,\sinh\alpha_{3}\cosh\alpha_{3} \left(\cosh^{4}\psi\, d\Omega_{4} \wedge d\psi\right. \nonumber\\
&&\qquad\qquad\left. -\frac{u}{2 (\mu^{4}_{3}\cosh^{2}{\alpha_{3}})^{2}}d\xi\wedge dx_{1}\wedge dx_{2}\wedge dx_{3}\wedge du \right).
\end{eqnarray}
Here we have kept the lowest order term of the expansions in each component of the metric and RR field because the next leading order term is suppressed by the factor $u^{2}/\mu^{4}_{3}$.
In order for the geometry to be geodesically complete, the compact direction ($\xi\sim\xi+2\pi\hat{R}$) should have a periodicity $\hat{R}=(\mu_{3}\cosh\alpha_{3})/2$. The geometry then becomes factorized into D$_{2}\times\mathbf{ R}^{3}\times$dS$_{4+1}$, where D$_{2}$ stands for a two-dimensional disk and the $5$-dimensional de Sitter space-time dS$_{4+1}$ has the cosmological constant $\Lambda_{4+1}=6/(\mu_{3}^{2}\,\cosh\alpha_{3})$.

\subsection{Exact Near-Bubble Solutions: Extremal Cases}

%Extremal cases have exact near-bubble solutions.

Though the near-bubble geometry involves de Sitter space-time, it should be understood only in the approximate sense. Meanwhile in the extremal case, one can expect an exact factorization of the near bubble geometry as the DWR version of the AdS$_{4+1}\times$ S$^{5}$. 

The extremal case can be obtained by limiting $\mu_{3}\rightarrow 0$ and $\alpha_{3}\rightarrow\infty$ with $Q_{3}=\mu_{3}^{4}\sinh\alpha_{3}\cosh\alpha_{3}$ kept finite. 
The resulting near bubble configuration is
\begin{eqnarray}\label{e15}
ds^{2}&=& \sqrt{Q_{3}} \left( \frac{1}{r^{2}}dr^{2}+ \frac{r^{2}}{Q_{3}} \left(d\xi^{2} +\sum^{3}_{i=1}dx_{i}^{2}\right)\right)+ \sqrt{Q_{3}}  \left(-d\psi^{2}+\cosh^{2}{\psi} d\Omega_{4}^{2}\right),\nonumber\\
\mathcal{F}_{(5)}&=&4iQ_{3}\left(\cosh^{4}{\psi}\,d\psi\wedge d\Omega_{4}- \frac{r^{3}}{Q^{2}_{3}} d\xi\wedge dx_{1}\wedge dx_{2}\wedge dx_{3}\wedge dr\right).
\end{eqnarray} 
The near-bubble geometry ($r^{4}\ll Q_{3}$) clearly includes $5$-dimensional de Sitter part (with the cosmological constant $\Lambda_{4+1}=6/\sqrt{Q_{3}}$). The first part in the above geometry actually describes a 5-dimensional hyperbolic space, whose Ricci tensor components are
\begin{equation}\label{e16}
R_{rr}=- \frac{4}{\sqrt{Q_{3}}}g_{rr},\qquad R_{\xi\xi}=- \frac{4}{\sqrt{Q_{3}}}g_{\xi\xi},\quad R_{ij}=- \frac{4}{\sqrt{Q_{3}}}g_{ij}\quad (i=1,\,2,\, 3).
\end{equation}  
The curvature scalar is given by $R=-20/\sqrt{Q_{3}}$. 

\subsection{Some Properties of the Bubbles}

The geometry is geodesically complete as far as $r\ge 0$ and there is no reason to make $\xi$ compact, being in contrast with the non-extremal case. However, one can take the direction $\xi$ compact to make a null bubble (a `bubble' in that the size of Kaluza-Klein compact direction vanishes at $r=0$ and `null' in that its wall follows a null line). The causal structures of the bubbles are drawn in Fig. 2. The space and time are restricted only to the shaded region, outside of which there is `nothing'.

\FIGURE{\epsfig{file=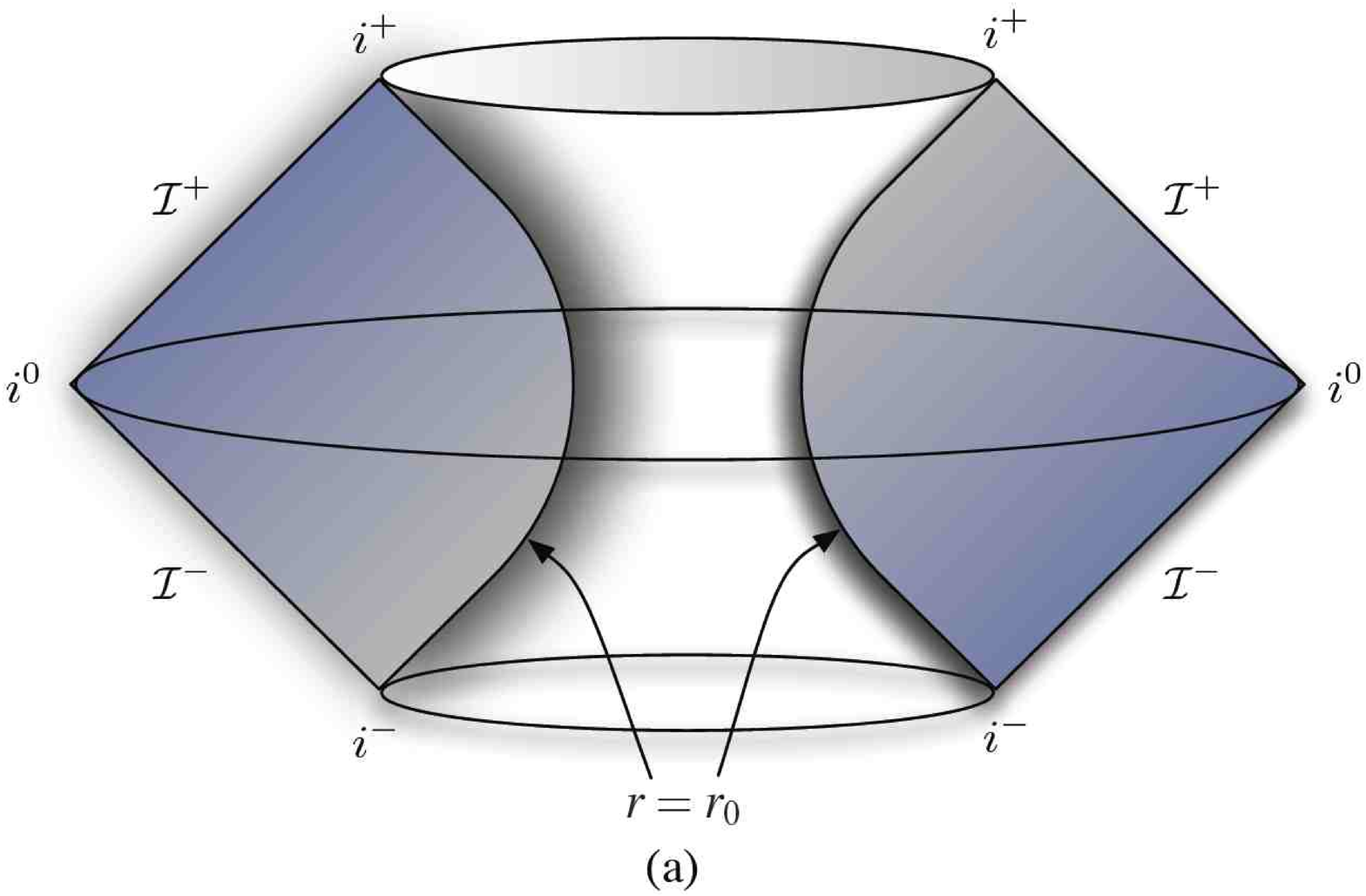,width=7cm} \epsfig{file=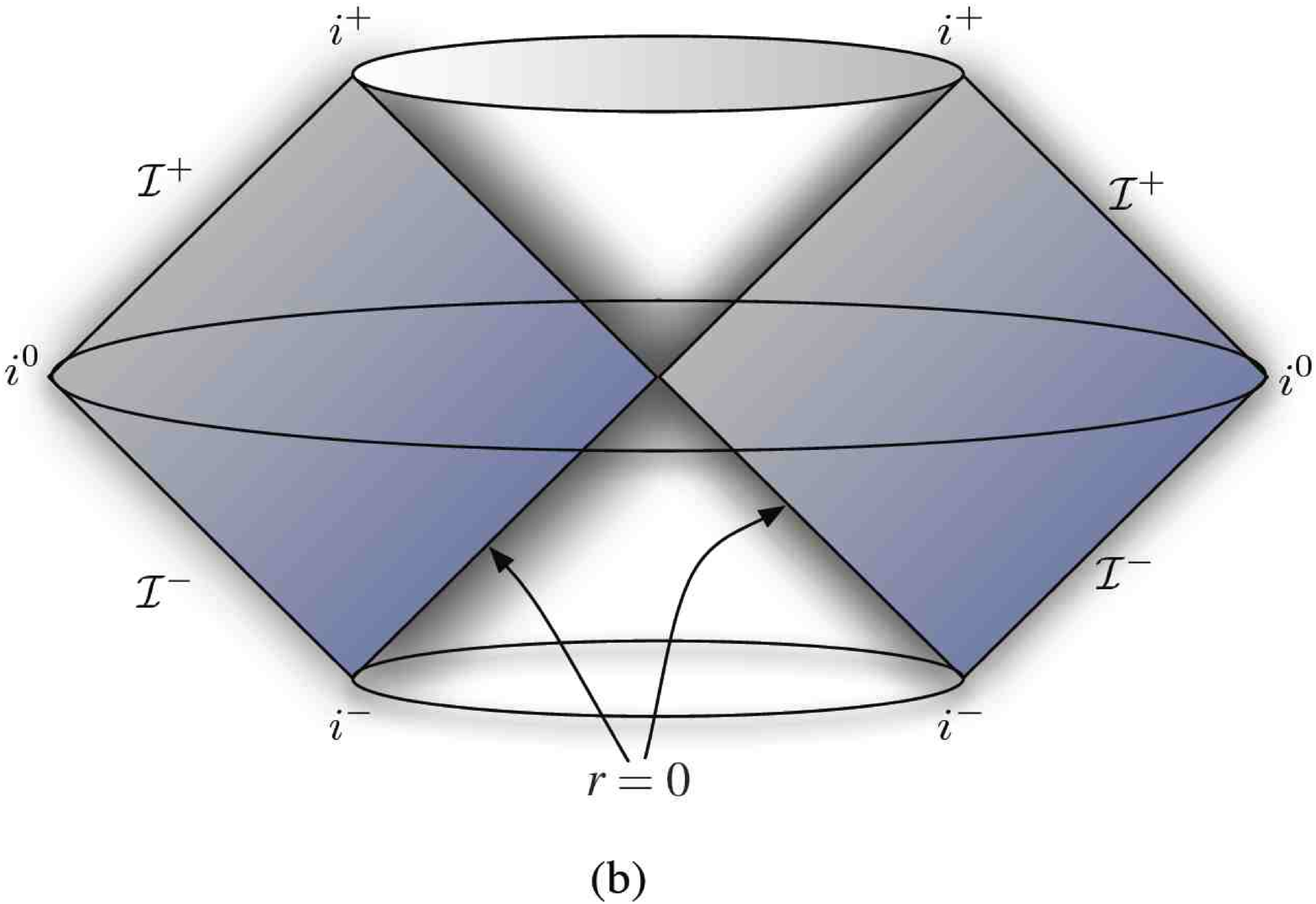,width=7cm} 
        \caption{Penrose diagrams for (a) the non-extremal bubble, and (b) the extremal bubble.}
	\label{figure2}}

%
%\begin{figure}[htbp]
%\begin{center}
%\includegraphics[width=10cm]{penrose.pdf}
%\caption{Penrose diagrams for (a) the non-extremal bubble, and (b) the extremal bubble.}
%\label{figure2}
%\end{center}
%\end{figure}

%\subsection{Extremal D$3$-bubbles as Hull's E$4$-branes in IIB}

We restrict our interest to the extremal case and try to identify the source that generates the bubble geometry. It carries the electric and magnetic RR flux, though imaginary. With the notion that its AdS cousin is the D$3$-branes with their worldvolume directions $(t,\,x_{1},\,x_{2},\,x_{3})$, it is natural to think of some $4$-dimensional object that extend to the directions $(\xi,\,x_{1},\,x_{2},\,x_{3})$. It is instanton-like in the sense that it is localized at the center in the transverse directions including the time $\psi$. 

Integrating the $5$-form field strength over the hyperbolic space that is transverse to de Sitter space-time, we obtain the magnetic flux
\begin{eqnarray}\label{e17}
\int_{ \text{H}_{5}} \mathcal{F}_{(5)}&=&-4iQ_{3}\int_{\text{H}_{5}}r^{-5}H^{-2}_{3} d\xi\wedge dx_{1}\wedge dx_{2}\wedge dx_{3}\wedge dr \nonumber\\
&=& -4iQ_{3} \text{Vol}_{4}\int^{\infty}_{0}dr \frac{r^{3}}{\left(r^{4}+Q_{3}\right)^{2}} \nonumber\\
&=&-i\text{Vol}_{4}
\end{eqnarray}  
and the electric flux,
\begin{eqnarray}\label{e18}
\int_{ \text{H}_{5}} *\mathcal{F}_{(5)}&=&4iQ_{3}\int_{\text{H}_{5}}* \left(\cosh^{4}{\psi}\,d\psi\wedge d\Omega_{4} \right) \nonumber\\ 
&=& 4iQ_{3}\int_{\text{H}_{5}}r^{-5}H^{-2}_{3} d\xi\wedge dx_{1}\wedge dx_{2}\wedge dx_{3}\wedge dr \nonumber\\
&=&i\text{Vol}_{4},
\end{eqnarray}
where $\text{Vol}_{4}$ is the coordinate volume of the anti-selfdual `instanton'. It is very interesting to see that both the electric and magnetic flux density are independent of the number $Q_{3}$ of D$3$-brane cousins. This is a universal feature common to most D/M-bubble configurations. In later part of this paper, we use this feature as one important clue to give the answer to the cosmological constant problem. 

One can take DWR on other D/M-brane configurations to get their corresponding D/M-bubble solutions. Since the procedure is straightforward, we defer the details to the appendix and only summarize the results in Table 1, where we tabulated the geometrical properties of the various D/M-bubbles. The second and the third columns summarize the near-bubble geometries of various non-extremal and extremal D/M-bubbles. In the non-extremal cases, the above factorized geometries are to be understood only in the approximate sense. The cosmological constants $\Lambda$'s are the same both in the extremal and in the non-extremal cases (the $4$-th column). The electric and the magnetic flux densities over the hyperbolic space are mostly fixed to $\pm i$ being irrespective of the number of source branes. The electric flux density of D$1$-D$5$-bubble is one exception and its value approaches to $i$ only as $Q_{1} \rightarrow Q_{5}$.

\begin{table}[htdp]
\begin{center}\begin{tabular}{||c||c|c|c|c|c||}\hline\hline \text{bubble type} & \text{non-extremal} & \text{extremal} & $\Lambda$ of dS& \text{electric}&magnetic\\\hline\hline \text{D}3 & D$_{2}\times \mathbf{ R}^{3}\times$dS$_{4+1}$& H$^{5}\times$dS$_{4+1}$ & $6/\sqrt{Q_{3}}$ &$ i$ &$-i$\\\hline \text{D}1-\text{D}5 &D$_{2}\times \mathbf{ R}^{5}\times$dS$_{2+1}$& H$^{7}\times$dS$_{2+1}$ & $1/\sqrt{Q_{1}Q_{5}}$ & $i\frac{\ln{\frac{Q_{1}}{Q_{5}}}}{(\frac{Q_{1}}{Q_{5}}-1)}$& $-i$\\\hline \text{M}2 & D$_{2}\times \mathbf{ R}^{2}\times$dS$_{6+1}$ & H$^{4}\times$dS$_{6+1}$ &$15/\sqrt[3]{Q_{T}}$  &  \text{none}&$i$\\\hline \text{M}5 & D$_{2}\times \mathbf{ R}^{5}\times$dS$_{3+1}$ & H$^{7}\times$dS$_{3+1}$ & $3/\sqrt[3]{Q^{2}_{F}}$ & $-i$ &\text{none}\\\hline\hline \end{tabular} \caption{Various D/M-bubbles}
\end{center}

\label{table1}
\end{table}

\subsection{Relation with Hull's E-branes}

With the temporal coordinate $\psi$ being transverse to the `world volume' directions, the bubble solutions could have some relation with Hull's Euclidean branes (E-branes) \cite{Hull:1998vg} or Gutperl and Strominger's spacelike branes (S-branes) \cite{Gutperle:2002ai}. (For more recent issues about E-branes or S-branes, see Ref. \cite{Durin:2005ts}.) Indeed D$3$-bubble solutions correspond to E$4$-brane solutions (compactified on a circle) in type IIB theory.  One can see this explicitly by adopting Rindler coordinates (\ref{rindler});
\begin{equation}\label{e19}
\tau=r\,\sinh{\psi},\qquad \rho=r\,\cosh{\psi}.
\end{equation}  
The D$3$-bubble solutions in these coordinates become
\begin{equation}\label{d3bubble}
ds^{2}=H_{3}^{ -\frac{1}{2}}(\tau,\,\rho) \left(d\xi^{2}+d\vec{x}^{2}  \right)+H_{3}^{ \frac{1}{2}}(\tau,\,\rho) \left( -d\tau^{2}+d\rho^{2}+\rho^{2} d\Omega^{2}_{4}\right),
\end{equation}
where the harmonic function is given by
\begin{equation}\label{e20}
H_{3}(r(\tau,\,\rho))=1+\frac{Q_{3}}{r^{4}}=1+ \frac{Q_{3}}{\left(\rho^{2}-\tau^{2} \right)^{2} }.
\end{equation}  
RR field strength becomes
\begin{eqnarray}\label{1imaginaryf5}
\mathcal{F}_{(5)}&=&4iQ_{3}\left(\frac{\rho^{4}}{\left(\rho^{2}-\tau^{2} \right)^{3} }\, d\Omega_{4} \wedge \left(\rho\,d\tau-\tau\,d\rho \right)\right.\nonumber\\
&&\quad\left. -\frac{1}{\left( \rho^{2}-\tau^{2}\right)^{3} H^{2}_{3}}d\xi\wedge dx_{1}\wedge dx_{2}\wedge dx_{3}\wedge \left(\rho\,d\rho-\tau\,d\tau \right) \right).
\end{eqnarray}  
The forms, (\ref{d3bubble}) and (\ref{e20}), of the solution are exactly the same as that of E$4$-branes except that the coordinate $\xi$ is compact. Therefore one can interpret the bubble solutions as E$4$-branes (and possibly S-branes) compactified on a circle S$^{1}$. 

Furthermore, the whole results suggest the equivalence between DWR and Hull's time-like T-duality. More specifically to say, DWR is the combination of T-dualities along one of the transverse directions and along the temporal direction;
\begin{equation}\label{e21}
\text{DWR}= \text{T}_{t}\cdot \text{T}_{\bot}.
\end{equation}    
In Hull's solution, RR field strength is real valued but it was instead embedded into a $*$-theory with unusual `wrong signed' kinetic terms for RR fields. One can recast the above solutions to those of $*$-theory by redefining RR fields as $\mathcal{F}_{(5)}\equiv iF_{(5)}$. We would like to emphasize that one virtue of DWR over the timelike T-duality is that the former can be applied to the solutions of M-theory, where T-duality is ambiguous.

\section{Multi-Bubbles\,:~ Swiss Cheese Universe}

\subsection{Maximal Supersymmetries near the Walls of Extremal `Bubbles'}
In order to make sense of multi-bubble solutions, we have to check whether there is any supersymmetry preserved by the bubble background. This issue is very confusing. On the one hand, there is a definite argument against the supersymmetry in de Sitter space-time. The argument is based on the fact that there is no positive conserved energy in de Sitter space-time. Roughly speaking, assuming any supersymmetry looks inconsistent with this fact. See Ref. \cite{Witten:2001kn} for details. On the other hand, we know the D/M-brane partners of the bubbles discussed in this paper preserve supersymmetries. Since DWR does not change the numbers of positive or negative signature in the metric, the spinor structure will be untouched. 

In the following, we show that although extremal `bubbles'\footnote{Supersymmetries are preserved only when the coordinate $\xi$ is not compact. In the strict sense, therefore, the supersymmetric solutions are not bubbles, but E-branes.} are time-dependent solutions\footnote{Precisely to say, time-dependence does not necessarily imply supersymmetry breaking. For some examples of time-dependent but supersymmetric configuration of D-branes, see Refs. \cite{Cho:2001ys}\cite{Cho:2002ga}\cite{Chen:2005mg}.} and have de Sitter regions, there are supersymmetries preserved. In fact, all of $32$ supersymmetries are preserved in the near-bubble regime of most D/M-bubbles (except D$1$-D$5$-bubble case where $16$ supersymmetries are preserved)  {\it as far as the coordinate $\xi$ is kept non-compact.} To be more specific, we will focus on M$5$-bubble case and obtain Killing spinors explicitly.    
The secret of this surprising result is mainly due to the imaginary valuedness of higher form fields. 

With all the details concerning vielbein components and spin connection components in appendix C,
the Killing spinor equations in eleven dimensional supergravity read as
\begin{equation}\label{e22}
0=\delta \psi_{\mu}=D_{\mu}\epsilon + \frac{1}{288} \left( \Gamma_{\mu}{}^{\nu\rho\sigma\kappa}-8\delta_{\mu}{}^{\nu}\Gamma^{\rho\sigma\kappa}\right) \epsilon \,\mathcal{F}_{\nu\rho\sigma\kappa},
\end{equation} 
where $\mathcal{F}_{(4)}$ denotes $4$-form field strength. The spinor $\epsilon$ is a $32$-component Majorana and 
\begin{equation}\label{e23}
\Gamma^{a_{1}a_{2}\cdots a_{n}}= \frac{1}{n!} \left(\Gamma^{a_{1}}\Gamma^{a_{2}}\cdots\Gamma^{a_{n}}+\sum( \text{anti-symmetric combinations}) \right). 
\end{equation}  
In the above equations, we used $11$-dimensional gamma matrices satisfying
\begin{equation}\label{e24}
\{\Gamma^{a},\,\Gamma^{b}\}=2\eta^{ab}.
\end{equation}

For the purpose of solving Killing spinor equations, it is convenient to work in the orthonormal frame, where $4$-form field strength is expanded in the case of S$^{4}$ or dS$_{3+1}$ as
\begin{eqnarray}\label{e25}
\mathcal{F}&=&-3Q\sin^{2}{\chi_{3}}\,d\Omega\wedge d\chi_{3}=-3Q^{- \frac{1}{3}}\,e^{7}\wedge e^{8}\wedge e^{9}\wedge e^{\natural} \nonumber\\
&=&-3iQ\sinh^{2}{\psi}\,d\Omega\wedge d\psi=-3iQ^{- \frac{1}{3}}\,\bar{e}^{7}\wedge \bar{e}^{8}\wedge \bar{e}^{9}\wedge \bar{e}^{\natural}.
\end{eqnarray}
In order to denote the cases of AdS$_{6+1}\times$ S$^{4}$ and H$^{7}\times$ dS$_{3+1}$ collectively, we represent the components $\eta^{00}\equiv\eta$ and $\eta^{\natural\natural}\equiv\kappa^{2}$ as
\begin{eqnarray}\label{e26}
\eta&=&-1 \qquad \kappa=1\qquad \text{(AdS$_{6+1}\times$ S$^{4}$)},\nonumber\\
\eta&=&1\qquad\,\,\,\,\,\kappa=i\qquad \text{(H$^{7}\times$ dS$_{3+1}$)}.
\end{eqnarray}
The Killing spinor equations are then recast in the orthonormal frame as
\begin{eqnarray}\label{kse}
0&=&\partial_{\mu}\epsilon + \frac{1}{4}\omega_{ab\mu}\Gamma^{ab}\epsilon+ \frac{1}{2^{5}\cdot 3^{2}} \left( e^{a}{}_{\mu}\Gamma_{a}{}^{bcde}-8e^{b}{}_{\mu}\Gamma^{cde}\right) \epsilon \,\mathcal{F}_{bcde} \nonumber\\
&\equiv&\partial_{\mu}\epsilon+ \frac{1}{2}\Omega_{\mu}\epsilon.
\end{eqnarray} 
The explicit forms of the matrices $\Omega_{\mu}$ are listed in Appendix C.4. Despite of their complicated expression, most components $\Omega_{\mu}$ are squared to $-1$ except $\Omega^{2}_{t}=\eta$, $\Omega^{2}_{\lambda}=\Omega^{2}_{\psi}=1$. One more property relevant in the computation of the Killing spinors is that $\Omega_{\mu}$'s of AdS$_{6+1}$/H$^{7}$ part and those of S$^{4}$/dS$_{3+1}$ part mutually commute. 

The final form of the Killing spinors satisfying all the above equations is
\begin{equation}\label{e27}
\epsilon=\Omega_{\text{AdS/H}}\cdot \Omega_{ \text{S/dS}}\,\cdot\epsilon_{0},
\end{equation}  
where 
\begin{eqnarray}\label{e28}
\Omega_{ \text{AdS/H}}&=& e^{ \frac{\theta_{4}}{2}\Gamma^{16}}e^{ -\frac{\theta_{3}}{2}\Gamma^{56}}e^{ -\frac{\theta_{2}}{2}\Gamma^{45}}e^{ -\frac{\theta_{1}}{2}\Gamma^{34}}e^{ -\frac{\varphi}{2}\Gamma^{23}}e^{ \frac{\kappa\lambda}{2}\Gamma^{1789\natural}}e^{ \frac{\kappa\eta t}{2}\Gamma^{0789\natural}}\nonumber\\
\Omega_{ \text{S/dS}}&=&\left\{\begin{array}{ll}e^{ \frac{\chi_{3}}{2}\Gamma^{789}}e^{ -\frac{\chi_{2}}{2}\Gamma^{9\natural}}e^{ -\frac{\chi_{1}}{2}\Gamma^{89}}e^{ -\frac{\phi}{2}\Gamma^{78}}&\quad (\text{S$^{4}$}) \\\\e^{\frac{1}{2}(i\psi+ \frac{\pi}{2})\Gamma^{789}}e^{ -\frac{i\chi_{2}}{2}\Gamma^{9\natural}}e^{ -\frac{\chi_{1}}{2}\Gamma^{89}}e^{ -\frac{\phi}{2}\Gamma^{78}} &\quad (\text{dS$_{3+1}$})\end{array}\right.
\end{eqnarray}
and $\epsilon_{0}$ is an arbitrary $32$-component constant Majorana spinor. Hence the whole supersymmetries are preserved for both AdS$_{6+1}\times$ S$^{4}$ and H$^{7}\times$ dS$_{3+1}$ cases. 

The secret of this supersymmetric de Sitter space-time lies in the imaginary field strength which enables the nice property of $(\Omega_{\mu})^{2}=\pm 1$, that is, they become coordinate independent once squared.
Lastly, we note one very interesting point about the periodicity of Killing spinors. Although the coordinate $t$ that was the temporal coordinate in AdS space-time, should be replaced by $\xi$ in the hyperbolic space and it should be kept {\it non-compact} to make Killing spinors sensible, because 
\begin{equation}\label{e29}
(\kappa\,\eta\,\Gamma^{0789\natural})^{2}=\eta=1.
\end{equation} 
On the contrary, the temporal coordinate $t$ should be quantizes as $t\sim t+2\pi n$ ($n\in\mathbf{Z}$) in AdS case, where we use the universal covering to avoid this unwanted feature.  

 The algebraic structure of de Sitter supersymmetry asserts that it can be nontrivally realized in the presence of ghosts (negative normed states, which come from the gauge fields with the wrong kinetic terms). One can extend de Sitter algebra to include supercharge operators for the extended superalgebra case. This was actually done by Lukierski and Nowicki \cite{Lukierski:1984it}, Pilch, Nieuwenhuizen, Sohnius \cite{Pilch:1984aw}, and Vasiliev \cite{Vasiliev:1986ye} separately. One crucial result about de Sitter supersymmetry is its algebraic structure different from anti-de Sitter supersymmetry counterpart. Conventional form of `$\sum_{i}\{Q_{i}, Q^{*}_{i}\} \sim H$' is \emph{not} valid in de Sitter case, where it is replaced by $\sum_{i}\{Q_{i}, Q^{*}_{i}\}=0$ ($i=1,\cdots,\mathcal{N}$). This means that de Sitter superalgebra cannot have any nontrival representation in Hilbert space (of positive normed states). However, this does not exclude the possibility of nontrivial representations in the presence of negative normed states, that is, ghosts. In Ref. \cite{Pilch:1984aw}, they explicitly construct an example of de Sitter supergravity model involving fields with the wrong sign kinetic term. See also Ref. \cite{Vasiliev:1986ye}. Hence in de Sitter case, supersymmetry can be nontrivially incorporated if we allow the fields with the wrong kinetic term. Note here that these ghost fields do \emph{not} imply that its energy can be unbounded below. In dS spacetime, there is \emph{no} notion of globally well-defined energy, as was discussed by Witten \cite{Witten:2001kn}. 

\subsection{Null Geodesic Lines in the Bubble Geometry}

In a realistic cosmological situation, one can expect that actually there are multiple bubbles. In case of two bubbles, both of which are expanding, the collision of bubbles might lead to the formation of singularity \cite{Horowitz:2002cx}. This sounds that the inhabitants near the bubble walls are destined to have a doomsday. 

However, if we live near the bubble wall, we will never witness the singularity formation. This means we cannot observe the bubbles swallow up us. For the purpose of seeing this, it is instructive to consider the null geodesic lines in the near-bubble coordinates $(\psi,\,r)$ and in the asymptotic Rindler coordinates $(\tau,\,\rho)$ respectively. In the former coordinates,
the null geodesic in the bubble background is determined by the equation
\begin{equation}\label{e30}
0=H (r)\left(dr^{2}-r^{2}d\psi^{2} \right). 
\end{equation}  
Here, $H(r)$ is a certain harmonic function specified by the brane type. In the near-bubble coordinates, the null geodesic lines passing through the point $(\psi_{i},\,r_{i})$ are represented as
\begin{equation}\label{e31}
\psi=\pm \ln{ \frac{r}{r_{i}}}+\psi_{i},
\end{equation}  
where the sign $+/-$ corresponds to the outgoing/ingoing null rays respectively. One thing to note is that it takes infinite coordinate time $(\psi \rightarrow \infty)$ for the null rays to reach or to come out of the bubble wall. 

The situation viewed from the asymptotic observer is different. In his coordinates $(\tau,\,\rho)=(r\sinh{\psi},\,r\cosh{\psi})$, the null geodesic lines are determined by
\begin{equation}\label{e32}
0=H(\tau,\,\rho) \left(d\rho^{2}-d\tau^{2} \right), 
\end{equation}  
and it takes finite coordinate time $(\tau<\infty)$ for the null rays to reach or to come out of the bubble wall. This property is illustrated in Fig. 3.

\FIGURE{\epsfig{file=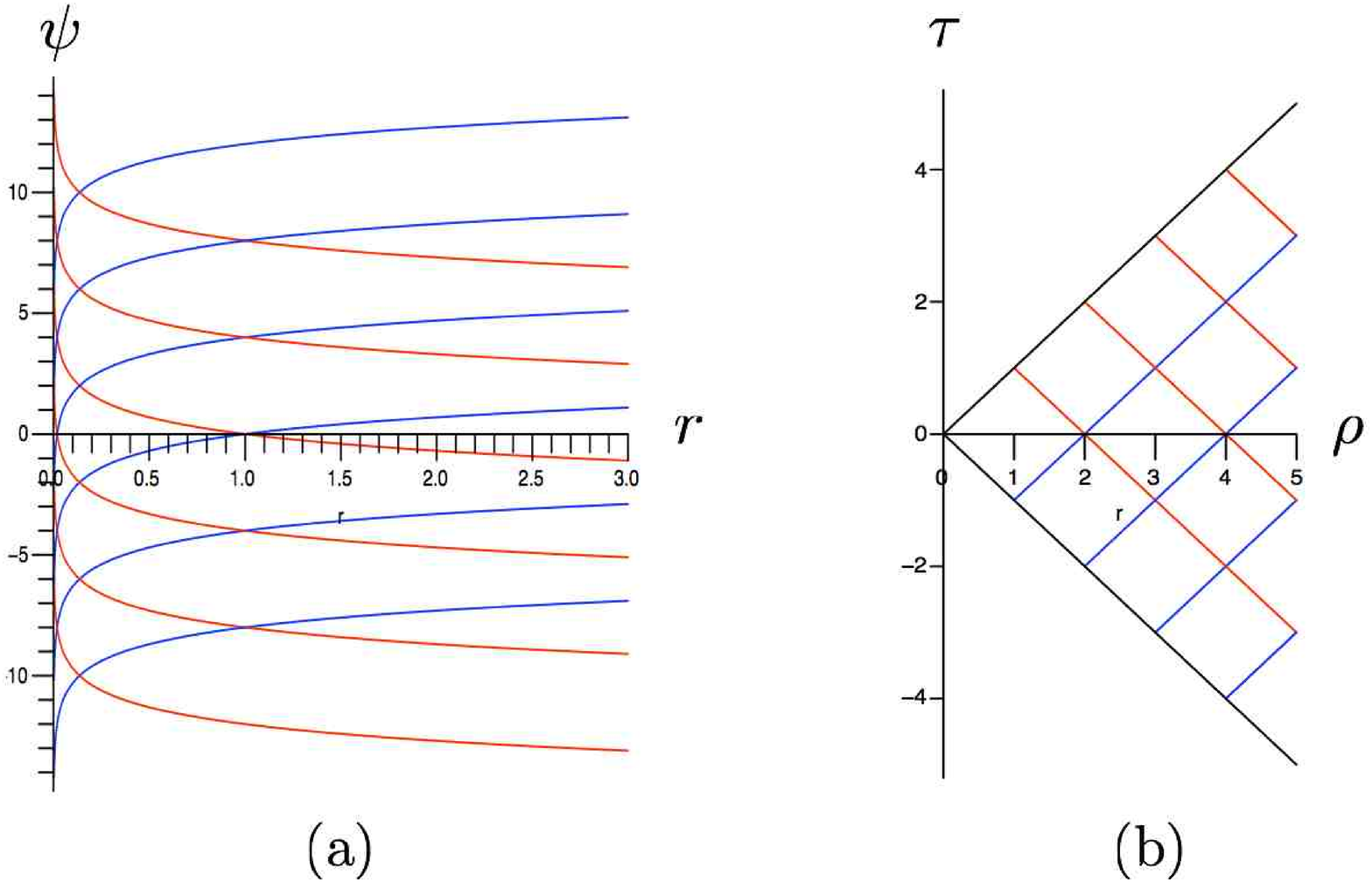,width=9cm} 
        \caption{Null geodesic lines (a) in the near-bubble coordinates $(\psi,\,r)$ and (b) in the asymptotic Rindler coordinates $(\tau,\,\rho)$. The red lines are incoming and the blue lines are outgoing.}
	\label{figure3}}

%\begin{figure}[htbp]
%\begin{center}
%\includegraphics[width=10cm]{geodesic.pdf}
%\caption{Null geodesic lines (a) in the near-bubble coordinates $(\psi,\,r)$ and (b) in the asymptotic Rindler coordinates $(\tau,\,\rho)$. The red lines are incoming and the blue lines are outgoing.}
%\label{figure3}
%\end{center}
%\end{figure}

Each intersection of the ingoing (red) null lines and the outgoing (blue) null lines composes a null cone. Especially in the near-bubble coordinates $(\psi,\,r)$, the null cones get pinched off and nothing can reach the wall in a finite coordinate time $\psi$. Actually this is related to the fact that $\psi$ itself is a good affine parameter and the bubble geometry is geodesically complete.

%\subsection{Non-constant Cosmological Constant}

\subsection{Bubble Landscape - Swiss Cheese Universe}

We will make use of supersymmetry to make sense of the multi-bubble solutions. However, in previous section, we showed that the bubble solutions keep most supersymmetries as with their D/M-brane cousins {only when the coordinate $\xi$ is non-compact.} On the other hand, we will see at later stage that it is necessary to compact all world volume directions (including $\xi$-direction) of Euclidean branes to make sense of lower-dimensional de Sitter gravity. 

Our strategy for the multi-bubble solutions is as follows. For the time being, we assume the coordinate $\xi$ to be non-compact, which will validate the `muti-bubble' solutions (though not bubbles in the strict sense). Upon compactification of $\xi$, the Killing spinors will no longer be single-valued on the periodic circle $\xi$ and \emph{the supersymmetry will be broken completely}. However, the bosonic sector will still satisfy the equations of motion which govern only the local behavior of the fields and do not care about the global property like the periodicity of the fermionic fields. In all,
starting from various multi-brane solutions, one can obtain their corresponding bubble cousins via the double Wick rotation. For simplicity, we focus on the extremal M$5$-brane case that results in various $4$-dimensional de Sitter space-times.

Generalizing the harmonic function to the multi-centered one, we get the multi-bubble solution;
\begin{equation}\label{e33}
H_{F} \Rightarrow 1+\sum_{a} \frac{Q_{a}}{|\vec{r}-\vec{r}_{a}|^{3} },
\end{equation}
   according to which $4$-form field strength becomes
\begin{equation}\label{e34}
\mathcal{F}_{(4)}=- 3i\sum_{a}\frac{Q_{a}r^{3}\vec{r}\cdot\left(\vec{r}-\vec{r}_{a} \right) }{|\vec{r}-\vec{r}_{a}|^{5}}\cosh^{3}\psi\,d\psi \wedge\,d\Omega_{3}\,.
\end{equation}  
If there is some region close to all the bubbles ($|\vec{r}-\vec{r}_{a}|^{3}\ll Q_{a}$), there, the metric looks like
\begin{eqnarray}\label{e35}
ds^{2}&=&\left(\sum_{a}\frac{Q_{a}}{|\vec{r}-\vec{r}_{a}|^{3}} \right)^{- \frac{1}{3}}\left( d\xi^{2}+\sum^{5}_{i=1}dx^{2}_{i} \right)  \nonumber\\
&&+\left(\sum_{a}\frac{Q_{a}}{|\vec{r}-\vec{r}_{a}|^{3}} \right)^{\frac{2}{3}}\left(dr^{2}-r^{2}\,d\psi^{2}+r^{2}\,\cosh^{2}\psi\,\, d\Omega_{3}^{2} \right)\,. 
\end{eqnarray}

\FIGURE{\epsfig{file=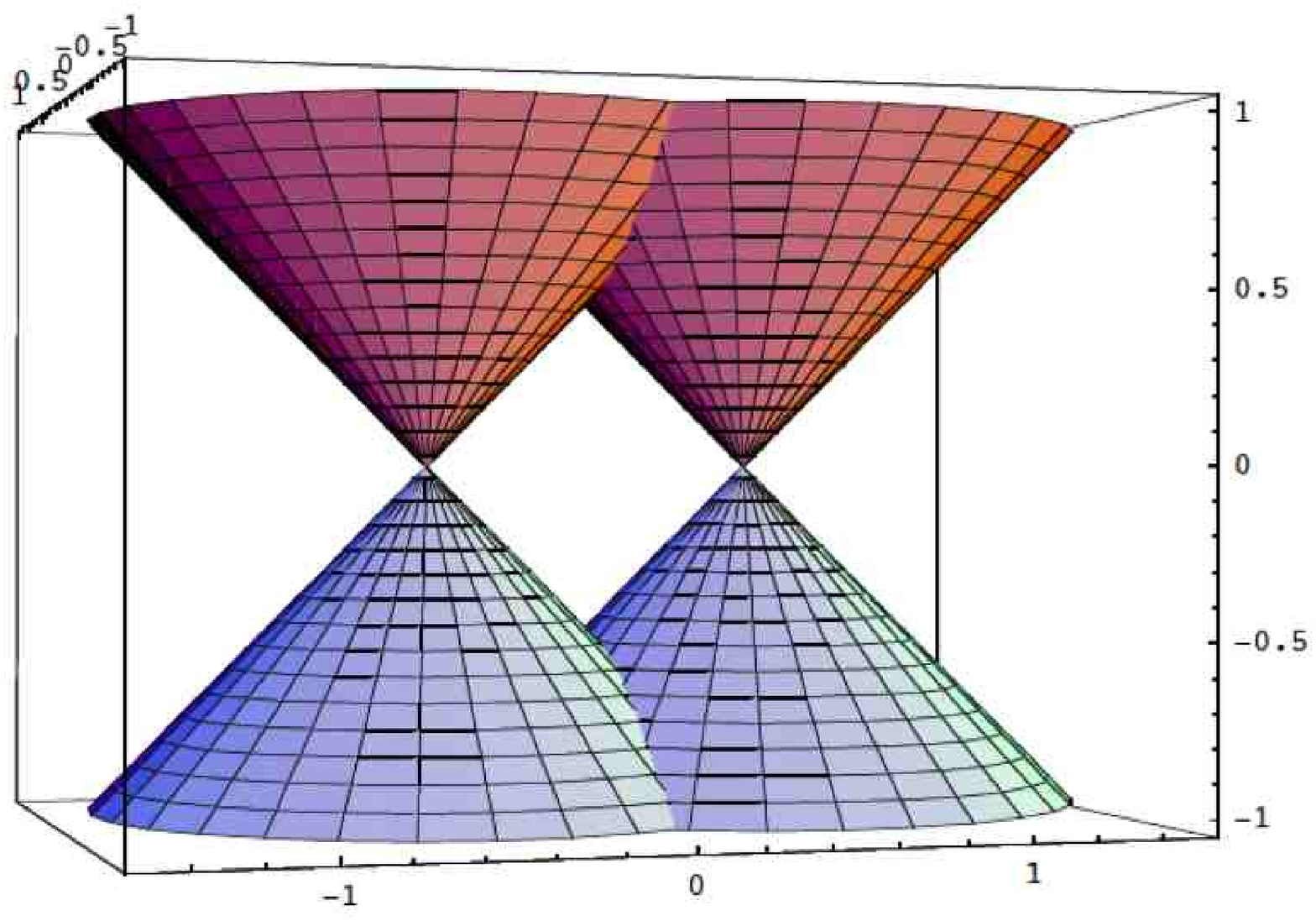,width=9cm} 
        \caption{Two colliding bubbles drawn in the asymptotic coordinates $(\tau,\,\vec{\rho})$. The cones bound the space-time outside, while there is `nothing' inside.}
	\label{figure4}}

%\begin{figure}[htbp]
%\begin{center}
%\includegraphics{twobubble.pdf}
%\caption{Two colliding bubbles drawn in the asymptotic coordinates $(\tau,\,\vec{\rho})$. The cones bound the space-time outside, while there is `nothing' inside.}
%\label{figure4}
%\end{center}
%\end{figure}
\noindent
Fig. 4 exhibits two colliding bubbles drawn in the asymptotic coordinates $(\tau,\,\vec{\rho})$. The geometry is defined only outside of both null cones. Near two bubbles, the geometry is almost factorized into de Sitter and the hyperbolic space but the effective cosmological constant varies from point to point. Some aspects of colliding Witten's bubble solutions were considered in Ref. \cite{Horowitz:2002cx}.

In order to see the resulting de Sitter space-time explicitly, let us be specific to the case of two-bubbles with equal charge just for simplicity. The harmonic function of two-bubbles at $\vec{r}_{1,2}=(\pm a,\,0,\,0,\,0)$ will be
\begin{equation}\label{e36}
H_{F}=1+ \frac{Q}{\left(r^{2}\sin^{2}{\theta_{1}}+ \left(r\cos{\theta_{1}-a} \right)^{2}  \right)^{ \frac{3}{2}} }+ \frac{Q}{\left(r^{2}\sin^{2}{\theta_{1}}+ \left(r\cos{\theta_{1}+a} \right)^{2}  \right)^{ \frac{3}{2}} }
\end{equation} 
In the region close to two bubbles, where the last two terms of the harmonic function become dominant over the constant `$1$', but with $a/r\ll 1$, that is in the region $a\ll r\ll \left(2Q \right)^{ \frac{1}{3}} $, the geometry is almost factorized into dS$_{3+1}\times$H$^{7}$;
\begin{eqnarray}\label{e37}
ds^{2}&=&\left( 2Q\right)^{- \frac{1}{3}}r \left( 1- \frac{1}{2}\left(5\cos^{2}{\theta_{1}}-1 \right)\frac{a^{2}}{r^{2}}+ \mathcal{O}\left( \frac{a^{4}}{r^{4}}\right) \right)\left( d\xi^{2}+\sum^{5}_{i=1}dx^{2}_{i} \right) \\
&&+\frac{ \left(2Q \right)^{ \frac{2}{3}}}{r^{2}}\left(1+\left(5\cos^{2}{\theta_{1}}-1 \right)\frac{a^{2}}{r^{2}}+ \mathcal{O}\left( \frac{a^{4}}{r^{4}}\right) \right)\left(dr^{2}-r^{2}\,d\psi^{2}+r^{2}\,\cosh^{2}\psi\,\, d\Omega_{3}^{2} \right)\,. \nonumber
\end{eqnarray} 
The effective cosmological constant in the region is $\Lambda\sim 3/(2Q)^{ \frac{2}{3}} $. If we restrict to the region very close to one of those two bubbles so that only one of the harmonic terms is relevant, then the corresponding cosmological constant will be increased to $3/Q^{ \frac{2}{3}}$. This could be a stringy realization of position-dependent cosmological constant. Such a concept is not new; quintessence \cite{Zlatev:1998tr} and k-essence \cite{Armendariz-Picon:2000ah} have such a feature. However we would like to emphasize that we need not introduce extra scalar fields to account for that. 

%\subsection{Bubble Landscape - Swiss Cheese Universe}

Though the multi-bubble geometry looks static in the above near-bubble coordinates, the bubbles are expanding in the light speed and collide with one another in the asymptotic Rindler coordinates $(\tau,\,\rho)=(r\sinh{\psi},\,r\cosh{\psi})$, in which the single bubble geometry is described as
\begin{eqnarray}\label{singlebubble}
ds^{2}&=&H_{F}^{- \frac{1}{3}}\left( d\xi^{2}+\sum^{5}_{i=1}dx^{2}_{i} \right) +H_{F}^{ \frac{2}{3}} \left(-d\tau^{2}+d\rho^{2}+\rho^{2}\, d\Omega_{3}^{2}\right). 
\end{eqnarray}
Here, the harmonic function $H_{F}$ is
\begin{eqnarray}\label{e38}
H_{F}=1+\frac{Q}{\left(-\tau^{2}+ \rho^{2}\right)^{\frac{3}{2}}}
=1+ \frac{Q}{\left( -\tau^{2}+\sum^{4}_{i=1}y^{2}_{i}\right)^{ \frac{3}{2}}}\,.
\end{eqnarray} 
In the last equality, we used the rectangular coordinates $y_{i}$ ($i=1,\cdots, 4$) in which the second part of the metric (\ref{singlebubble}) is written as $H^{2/3}_{F}(-d\tau^{2}+\sum^{4}_{i=1}\,dy^{2}_{i})$. We note here that the geometry is restricted to the region exterior to the null cone $\rho^{2}=\tau^{2}$ because $\rho^{2}-\tau^{2}=r^{2}>0$. Especially if $\xi$ is compact, the asymptotic region where $H_{F}\simeq 1$, is a flat $(9+1)$-dimensional space-time compactified on a circle. At $r^{2}=\rho^{2}-\tau^{2}=0$, the compact dimension pinches off because any finite period of $\xi$ is suppressed by the factor $H^{-1/3}_{F}$. Therefore the whole geometry exterior to the null cone is geodesically complete space-time without any singularity and describes the null shock wave of a `bubble of nothing' collapsing and re-expanding. 

The harmonic function corresponding to the multi-bubble solution will be
\begin{equation}\label{e39}
H_{F} \Rightarrow 1+\sum_{a} \frac{Q_{a}}{\left(-\left( \tau-\tau_{a}\right)^{2}+\left( \vec{y}-\vec{y}_{a}\right)^{2} \right)^{ \frac{3}{2}} }\,.
\end{equation}
The bubbles are located at $(\tau_{a},\,\vec{y}_{a})$ and are separated from one another by the spacelike distances because $-\Delta \tau^{2}+\Delta \vec{y}\,^{2}=\Delta \vec{r}\,^{2}>0$. In the same vein, the geometry is restricted to the region exterior to all the bubbles.

\FIGURE{\epsfig{file=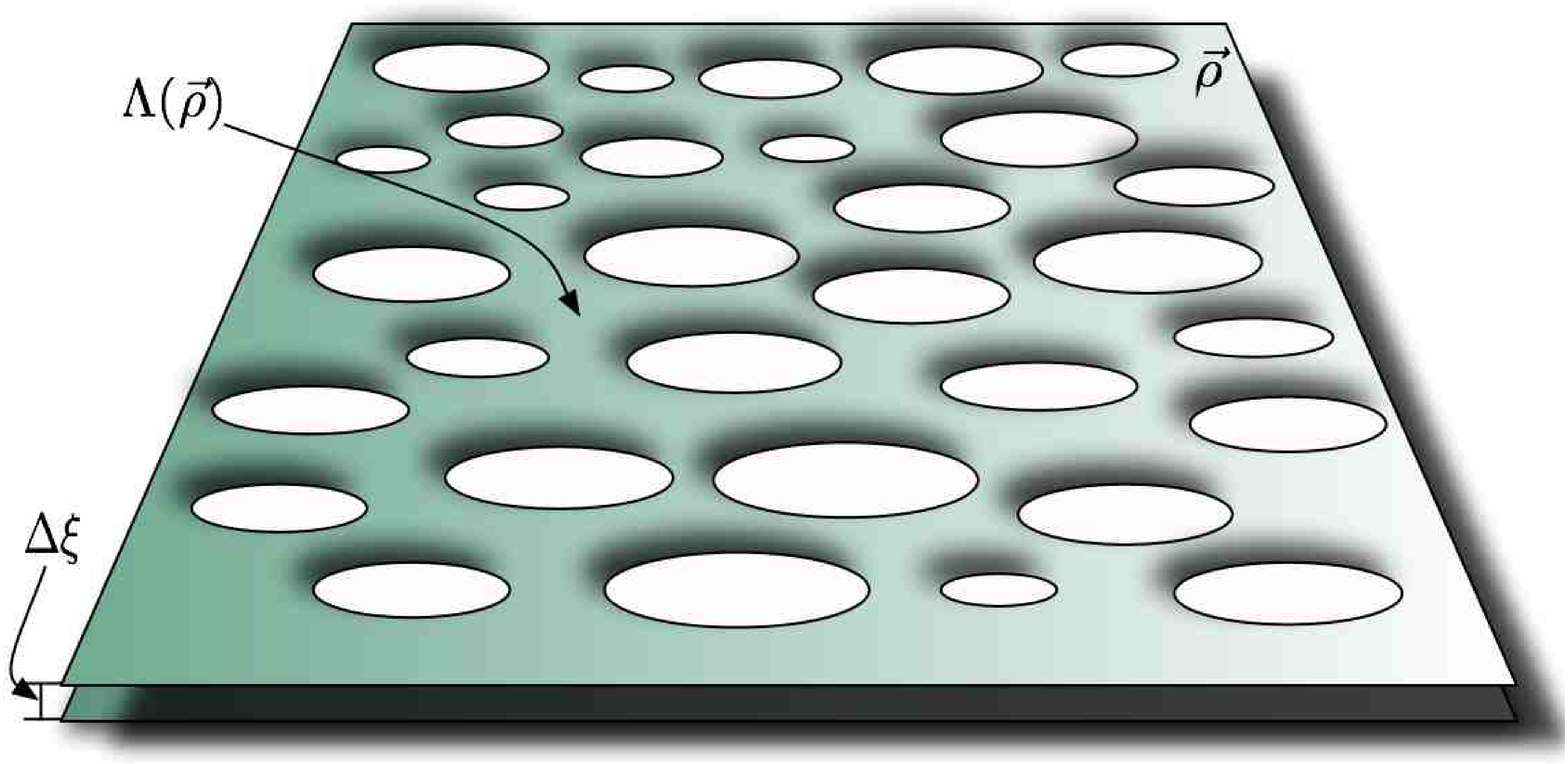,width=9cm} 
        \caption{Swiss-cheese-like geometry of a multi-bubble solution}
	\label{figure5}}

%\begin{figure}[htbp]
%\begin{center}
%\includegraphics[width=12cm]{bubbleland.pdf}
%\caption{Swiss-cheese-like geometry of a multi-bubble solution}
%\label{figure5}
%\end{center}
%\end{figure}

Fig. 5 visualizes Swiss-cheese-like geometry of a multi-bubble solution. The bubbles are expanding at the light speed but no observer can see them swallow up the whole space-time completely because all the observers will be left, being squeezed by the expanding bubbles, to the near-bubble region where time coordinate $\psi$ is the affine parameter. Depending on the position in $\vec{\rho}\in \mathbf{R}^{4}$, the effective cosmological constant varies. Notice that there is a `Universe' corresponding to each bubble and the multi-bubble background gives rise to the multi-verse picture. Some of the earlier ideas of multi-verse can be found, for example in Refs.  \cite{Linde:1988ws}\cite{Nam:1999tz}. Varying cosmological constant in the multi-verse is reminiscent of more recent idea of landscape \cite{Susskind:2003kw}.

\section{Semi-classical Instability of de Sitter Space-times}
\subsection{The Creation of Spherical M$2$ Branes in M$5$-Bubble Background}
The only price we have to pay for de Sitter solutions in string/M theories was to allow the imaginary field strengths. In this section, we interpret this disturbing situation as one mechanism to temper the acceleration of the universe so that it lead to de Sitter space-time with almost, but not exactly vanishing cosmological constant like our present universe.

Before going into details, we note that the imaginary valued term in the quantum effective action indicates the instability of the vacuum \cite{Schwinger:1951nm}. This statement is valid even in the classical action. Recall that Schr\"{o}dinger equation tells us that the imaginary potential term in the classical action leads to the source term in the continuity equation for the probability density and the probability flux density;
\begin{eqnarray}\label{e40}
\frac{\partial}{\partial t}\rho(t,\,\vec{x})+\nabla\cdot \vec{j} 
&=&\frac{\partial}{\partial t}\left(\psi^{*}(t,\,\vec{x})\psi(t,\,\vec{x})\right) +\nabla\cdot \left( \frac{\hbar}{m} \text{Im}\left(\psi^{*}(t,\,\vec{x})\nabla \psi(t,\,\vec{x})\right) \right)\nonumber\\
& =& \frac{i}{\hbar} \left(V^{*}(\vec{x})-V(\vec{x}) \right)\left(\psi^{*}(t,\,\vec{x})\psi(t,\,\vec{x}) \right). 
\end{eqnarray}
Therefore the imaginary potential term in the {\em classical action} indicates the violation of the unitarity, in other words, the creation of probability, at the {\em quantum} level.    

Though the supergravity action is invariant under DWR, therefore real valued for the specific bubble  solutions discussed so far, there might be some imaginary valued term being involved when probe D/M-branes are on their ways in these backgrounds. If so, we have to worry about the instability of the backgrounds themselves. Indeed the backgrounds are unstable at the semi-classical level. 

To be more explicit, let us focus on the M$5$-bubble geometry, in which de Sitter part of the near-bubble metric is given by
\begin{equation}\label{e41}
ds^{2}_{3+1}=Q^{ \frac{2}{3}} \left(-d\psi^{2}+\cosh^{2}{\psi} \left(d\theta^{2}_{1}+\sin^{2}{\theta_{1}}d\Omega^{2}_{2} \right)  \right). 
\end{equation}  
Let us assume a spherical membrane composed of multi-sheets at constant $\theta_{1}$, then it will couple to the imaginary $4$-form field strength via
\begin{equation}\label{e42}
q\int C^{(3)}=q\int d\psi \cosh^{3}{\psi}\wedge A^{(2)},
\end{equation}    
where $q$ is proportional to the number of the membrane sheets and its sign denotes the orientation of the membrane. The part $A^{(2)}$ composing the gauge connection has been introduced just for the  notational convenience and is given by
\begin{equation}\label{e43}
A^{(2)}=\left\{\begin{array}{ll}A^{(2)}_{N}=\frac{-3iQ}{2}\left(\frac{\sin{2\theta_{1}}}{2}-\theta_{1} \right)d\Omega_{2}  &\quad (0\le \theta_{1}< \pi), \\
&\\
A^{(2)}_{S}=\frac{-3iQ}{2}\left(\frac{\sin{2\theta_{1}}}{2}-\theta_{1}+\pi \right)d\Omega_{2} &\quad (0< \theta_{1}\le \pi).\end{array}\right.
\end{equation} 
In order to make the gauge fields well-defined, we introduced different gauges field in the northern and the southern hemi-3-sphere. In spherical coordinates, every connection form field concerning the azimuthal direction is ill-defined at the poles unless its value vanishes there. This is in analogy with the case of Dirac monopole. The difference of those two gauge fields, at any point of constant $\theta_{1}$ different from $ 0$ or $\pi$, is an exact form involving
\begin{equation}\label{e44}
A^{(2)}_{N}-A^{(2)}_{S}=\frac{3\pi iQ}{2}d\Omega_{2}.
\end{equation}    

The spherical membrane embedded into the spatial section S$^{3}$ of de Sitter bounds two $3$-balls. The field strengths inside both balls induce different gauge fields via Stokes theorem on their common boundary, the spherical membrane.  In Fig. 6, we draw the relative values of the gauge fields $A^{(2)}_{N}$ (from the north $3$-ball) and $A^{(2)}_{S}$ (from the south $3$-ball). 

\FIGURE{\epsfig{file=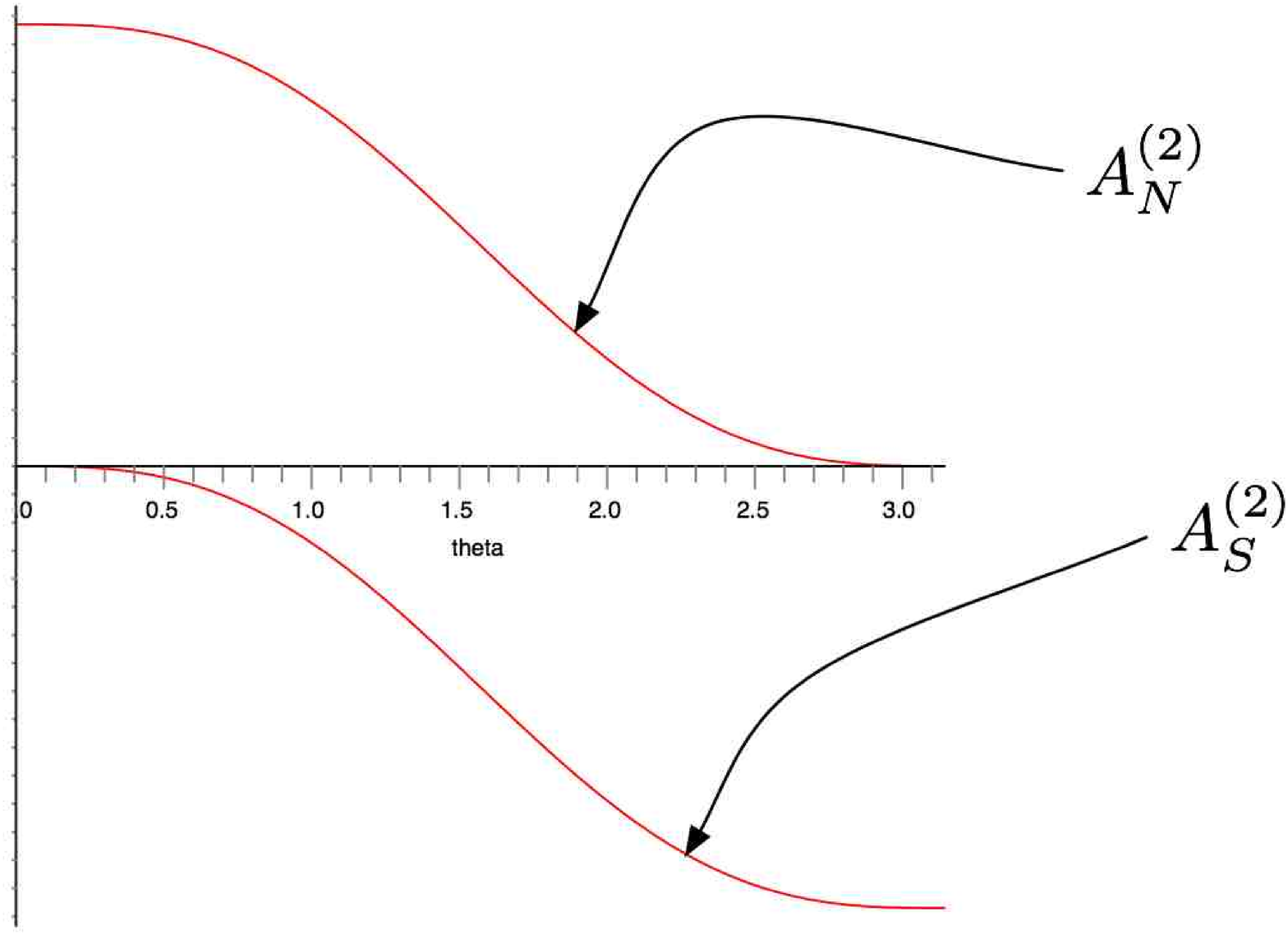,width=10cm} 
        \caption{The relative values of the gauge fields $A^{(2)}_{N}$ (from the north $3$-ball) and $A^{(2)}_{S}$ (from the south $3$-ball)}
	\label{figure6}}

%\begin{figure}[htbp]
%\begin{center}
%\includegraphics[width=3in]{connection.pdf} 
%\caption{The relative values of the gauge fields $A^{(2)}_{N}$ (from the north $3$-ball) and $A^{(2)}_{S}$ (from the south $3$-ball)}
%\label{figure6}
%\end{center}
%\end{figure}

Despite of the analogy with Dirac monopole, there are some differences between them. First, the imaginary interaction term causes some instability of the vacuum (pure de Sitter space-time). When the above interaction term gets exponentiated with extra quantum factor $i/\hbar$, it gives a decaying factor, rather than the conventional oscillatory phase. This could be understood as the instability in the vacuum-to-vacuum transition in the semi-classical sense.  Second, the instability depends on whether the spherical membrane couples to the connection involving $A^{(2)}_{N}$ or $A^{(2)}_{S}$. Recall that the gauge choice does not matter in the monopole case due to Dirac quantization condition of the electric and magnetic charges. In the case at hand, the electric flux over the whole de Sitter space-time is not a conserved quantity\footnote{The conserved electric flux is over the transverse $7$-hyperboloid instead.} and what is worse is that such a condition as Dirac quantization is not helpful in removing the gauge ambiguity.  

In order to fix up this ambiguity, we have to find a gauge invariant interaction of the spherical membrane with the imaginary background field. As was alluded earlier, the spherical membrane being embedded into the spatial $3$-sphere, bounds two regions of $3$-balls. It is reasonable to think that the spherical membrane couples to the background gauge fields \emph{from both $3$-balls in a gauge invariant way}:
\begin{eqnarray}\label{e45}
q\int d\psi\cosh^{3}\psi\int_{S^{2}} \left(A^{(2)}_{N}-A^{(2)}_{S} \right) &=& q\int d\psi\cosh^{3}\psi\int_{S^{2}} \frac{3\pi iQ}{2}d\Omega_{2}\nonumber\\
&=&6\pi^{2}iqQ\int d\psi\cosh^{3}\psi\,.
\end{eqnarray}
The relative different signs of the gauge potentials in the first expression are due to different orientations of the north and the south $3$-balls which the spherical membrane bounds.

This imaginary interaction term represents the instability of pure de Sitter vacuum against the creation of spherical membranes over its $3$-spherical space. This is similar to Schwinger pair production procedure of charged particles in a strong electric field. Recall that the one-loop diagram of charged particles in the background of the strong electric field results in the pure imaginary term in the effective action, which implies the instability of the corresponding vacuum and signals the pair creation of the charged particles \cite{Schwinger:1951nm}. In the above, we get a pure imaginary term in the interaction of a spherical membrane (the composite of $q$ M$2$-branes and $q$ anti-M$2$-branes) with de Sitter background. When $qQ>0$, one can interpret this term in the semi-classical sense as the instability of the de Sitter vacuum and signals the creation of spherical membranes with the rate, 
\begin{equation}\label{e46}
2\, \text{Im}\, \mathcal{L}=6\pi^{2}qQ\cosh^{3}{\psi},
\end{equation} 
per unit coordinate volume and per unit coordinate time. 
In all, M5-bubble de Sitter vacua are unstable against the creation of spherical membranes (See Fig.~7(a)). 

Some might be afraid that thus created spherical membranes will be annihilated by spherical membranes of the opposite charge. However in our case, one should note one big difference from the case of particle pair creation. In the particle case, the particle and the anti-particle are created in pair as two distinct objects in an unstable background. Meanwhile, a single compact brane (extended object) can be considered as the composite of one brane and its anti-brane partner. Therefore the precise analogy with Schwinger's creation of a particle pair is the creation of a single compact brane. 

Despite this fact, one can conceive of another kind of spherical branes with the opposite orientation (i.e., with their inside out, or with the opposite sign of $q$). However, the spherical membrane with the opposite orientation, if ever it was present, will be annihilated by the same background because of the opposite sign of $q$ in Eq. (\ref{e46}). Therefore in a given background of bubbles, only one kind of spherical branes are created and the branes of the opposite sign of $q$ will be suppressed. From the beginning, there are not spherical membranes with the opposite orientation, to be paired off with the created spherical membranes. 

The creation rate of the spherical membranes is the same over all the latitudes (Fig.~7(b)). The spherical symmetry of the spatial $3$-sphere of de Sitter and the fact that the above result does not depend on the latitudinal position of the membranes imply the random creation of the spherical membranes of random sizes over the spatial $3$-sphere (Fig.~7(c)).    

\FIGURE{\epsfig{file=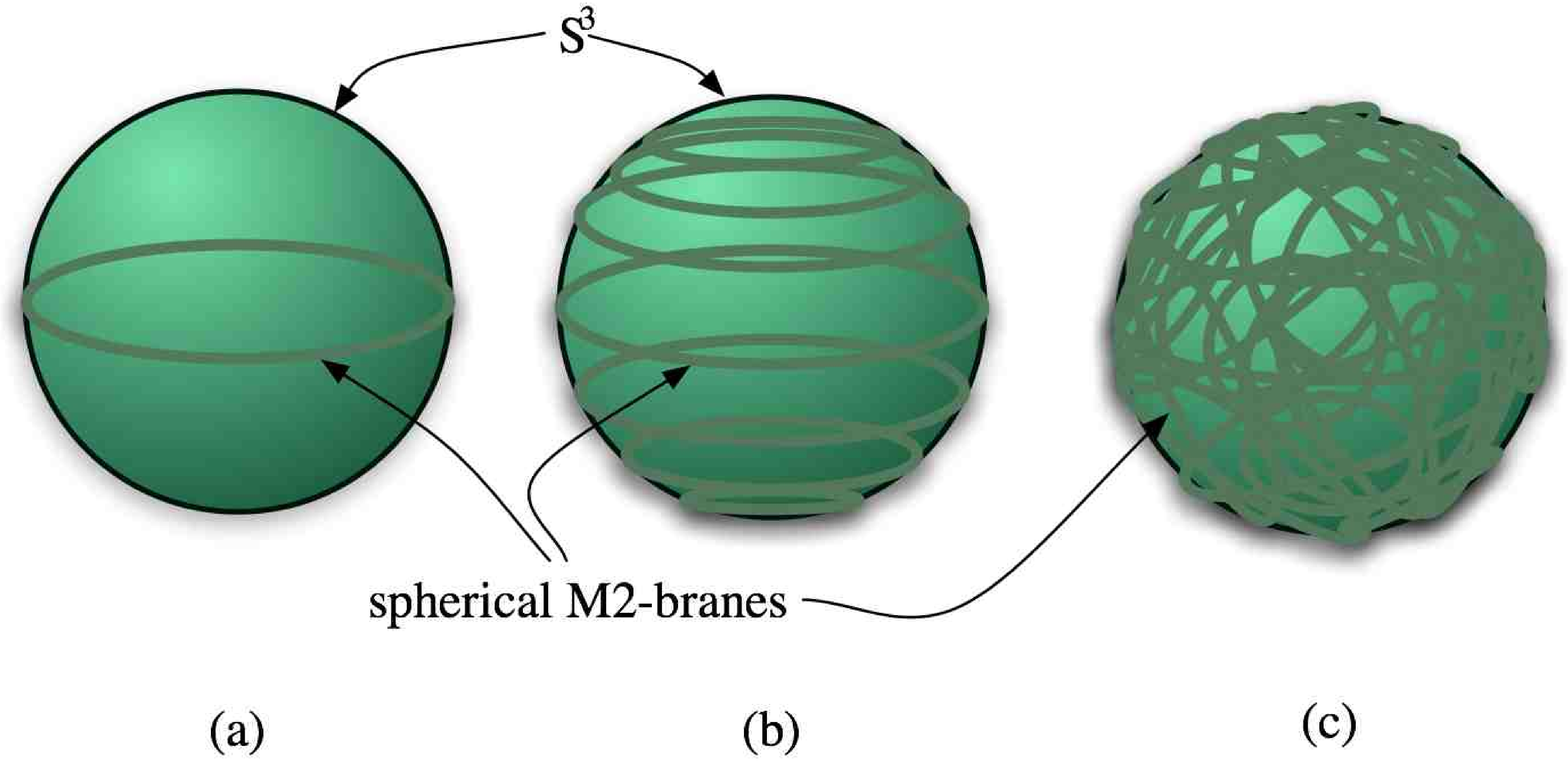,width=12cm} 
        \caption{The condensation of spherical membranes over S$^{3}$, the spatial part of de Sitter.}
	\label{figure7}}

%\begin{figure}[htbp]
%\begin{center}
%\includegraphics[width=5in]{m2wrapping1.pdf}
%\caption{The condensation of spherical membranes over S$^{3}$, the spatial part of de Sitter.}
%\label{figure7}
%\end{center}
%\end{figure}
%\noindent

\subsection{Spherical M$2$ Branes and the Cosmological Constant Problem}\label{4.2}
Let us consider the physical implication of the condensation of the spherical membranes. Though it is difficult to compute the back reaction of the condensation in the present setup, one can think of its qualitative consequence. The condensation of the spherical membranes provides us with a natural mechanism to explain the cosmological constant problem, that is, why the cosmological constant is almonst vanishing but not comletely. The tension of the created spherical membranes will squeeze the accelerating spatial $3$-sphere of de Sitter space-time \`{a} la Brandenberger-Vafa mechanism \cite{Brandenberger:1988aj}, which results in the reduction of the cosmological constant. (See also \cite{Tseytlin:1991xk} and \cite{Easther:2002qk} for detailed analysis.) This is similar to the reheating procedure where matters nucleate as the inflation stops. Here, the condensation of the spherical membranes extracts energy from de Sitter background, which lowers the cosmological constant\footnote{Some of the early ideas of neutralizing the cosmological constant were considered in Refs. \cite{Abbott:1984qf} \cite{Brown:1987dd} \cite{Brown:1988kg} \cite{Bousso:2000xa}.}. The created spherical membranes in this manner could provide the seed of matters.

This procedure cannot exhaust the cosmological constant completely. Recall that the Schwinger pair creation process does not violate the charge conservation. Despite of the condensation of the spherical membranes (with vanishing net M$2$ charge), the electric flux density `$i$' over the spatial $7$-hyperboloid should be invariant. This implies that the resultant geometry of the condensation of the spherical membranes cannot be completely flat Minkowski space-time where the imaginary electric flux is absent. Though the cosmological constant approaches to `$0$' from above, it cannot vanish completely. This could be one explanation of the cosmological constant problem.

\subsection{Any Pathology of the Ghosts?}

The conventional pathologies, which have been discussed in the theories involving the ghost fields\footnote{Here, the ghost fields mean the ones whose kinetic terms are wrong-signed (or equivalently the imaginary valued fields), thus result in the negative energy and the negative normed state at the quantum level.}, are as follow \cite{Hawking:2001yt}:
Largely we could have classical or quantum instabilities. What makes them serious is their runaway behavior, that is, exponential growing. For example, in the classical sense, the energy can flow from the ghost fields to the ordinary matter fields forever. Meanwhile in the quantum theory, the probability can flow from the ghost fields to the ordinary matters forever. Let us check whether the ghost form fields in the bubble solutions, discussed in this paper, lead to these kinds of runaway behaviors. 

In this paper, we show that this disastrous situation does not happen at least in the near-bubble wall. We confine our interests to the near-bubble wall region only and discuss the substantial difficulties involved in the full ten- or eleven-dimensional analysis in the later part.
Since the near-bubble geometry itself is an exact solution, it is yet meaningful to check the pathological runaway of the instability caused by the ghosts living near de Sitter bubble wall. 

In the near-bubble wall region, one has to make clearer the meaning of the classical instability. The factorization of $\mbox{dS} \times \mbox{H}$ enables us to pursue the argument just in dS separately. However, the energy is not globally well-defined in de Sitter spacetime, due to which the instability argument via the energy is ambiguous.
Instead, we have to check whether the ghost fields increase their density to grow the cosmological constant of de Sitter spacetime. This is not likely to be the case. Let us focus on M$5$-bubble case, just for simplicity. 

First, consider the local fluctuations of the cosmological `constant' over de Sitter. They will be sourced by the generic matter satisfying the equation of state $w=-1$. (We will see later that the ghost fields discussed in this paper are of this type satisfying the condition.) The equation of state renders those fluctuations physically trivial, that is, it leads to a rigidity theorem which excludes local excitations in the cosmological constant as physical excitations \cite{Ginsparg:1982rs}. 

Therefore, we focus on the solutions respecting de Sitter isometry ($SO(7,1)\times SO(4,1)$ pulled back onto $\mbox{H}^7 \times \mbox{dS}_{3+1}$ for M5 bubble case). The four-form fields concerning the ghosts compatible with the isometry will take the full volume forms (of $\mbox{dS}_{3+1}$), thus become dynamically trivial. In other words, de Sitter isometry will determine the field strength to be just a constant and the corresponding equation of motion, 
\begin{eqnarray}\label{}
\partial_{\mu}\mathcal{F}^{\mu\nu \rho\sigma}- \frac{\epsilon^{\mu_{1}\mu_{2}\cdots\mu_{8}\nu\rho\sigma}}{6\cdot 4!\cdot 3!}\left(3\partial_{\mu_{1}}\mathcal{F}_{\mu_{2}\mu_{3}\mu_{4}\mu_{5}}C_{\mu_{6}\mu_{7}\mu_{8}}+\mathcal{F}_{\mu_{1}\mu_{2}\mu_{3}\mu_{4}}\mathcal{F}_{\mu_{5}\mu_{6}\mu_{7}\mu_{8}} \right) &=&0.
\end{eqnarray}
will be satisfied automatically. The physical meaning of the constant can be read from Einstein equation,
\begin{equation}\label{11eom}
R_{\mu\nu}- \frac{1}{2}g_{ \mu\nu}R= \frac{1}{2\cdot 3!}\mathcal{F}_{\mu\rho\sigma\kappa}\mathcal{F}_{\nu}{}^{\rho\sigma\kappa}- \frac{g_{ \mu\nu}}{2\cdot 2\cdot 4!}\mathcal{F}_{\lambda\rho\sigma\kappa}\mathcal{F}^{\lambda\rho\sigma\kappa},
\end{equation}  
and is nothing but the cosmological constant. Therefore different value of the ghost corresponds to different choice of the vacuum, thus different `superselection sector', which is not allowed classically to change. 
        
Let us turn to the possibility of quantum runaway instability. In the near bubble region, we do not have any tool to deal with this issue because the quantum field theory in de Sitter spacetime is not well established yet. However one can still resort to some semiclassical way. 

One possibility is the `gravitational' instanton solution (caused by the ghosts) in de Sitter spacetime. The strict gravitational instanton solutions describing the creation of black holes in de Sitter spacetime have been discussed much so far \cite{Abbott:1984qf} \cite{Ginsparg:1982rs}. Since Hawking temperature of these black holes are higher than the temperature of de Sitter spactime, they will be evaporated away and the geometry itself will remain stable against this possibility. 

On the other hand, in our bubble solutions, the ghost fields might do some role for the possible quantum instability and it actually turns out to be the case. However, the instability does not lead to the disastrous runaway. Instead of increasing the cosmological constant, the ghost fields destabilize it to lower values. As was discussed in previous subsection \ref{4.2}, the ghost field creates the spherical membranes randomly over de Sitter, due to which the cosmological constant cascades to lower value but never to zero.

Despite all these good signatures in the near-bubble regime, the argument is clearly far from being complete. There might be some pathology involved in the full solutions embedded into ten- or eleven-dimensions. Though the geometry is asymptotically flat, so the energy-momentum is well defined, there are several difficulties concerning the analysis about  the eternal energy flow from the ghost fields to the geometry. 

For example, the temporal direction $\partial/\partial\tau$ of the asymptotic flat region is not Killing as we see in Eq. (\ref{singlebubble}), therefore the net energy of the gravitational excitation and the excitation of the ghost fields is not conserved. (Even we do not know whether the analysis via the energy in this case is sensible.) 

Moreover, in distinction to the near bubble wall case, the radial fluctuation of the ghost field cannot be excluded just from the isometry view point. The radial fluctuation of the ghost field could destabilize the whole solution. (However, there is still a bit of hope on this. Even in the worst case, the story in this paper could be relevant at least in the early era of the Universe. Being lacking of the tool beyond Planck scale, we inevitably have to introduce the trans-Planckian cutoff in order to avoid the initial singularity; for example the minimal length via some generalized uncertainty principle of the trans-Plackian physics or the fundamental scale of the string theory. If this kind of cutoff is one of the essential ingredients of the unknown trans-Planckinan physics, it will control the possible runaway behavior of the energy effectively. For the roles of the trans-Planckian cut-off taming the ultra-violet divergences, see Refs. \cite{DeWitt:1964yh} \cite{Padmanabhan:1988jp} \cite{Hassan:2002qk}.)  We defer this tough analysis elsewhere \cite{cho3}. 

\section{Bubble Nucleation}

\subsection{Bubbles vs. Instability of Kaluza-Klein Vacua}
In this subsection, we look into the origin of the bubble nucleation. Let us first see the possibility of the bubble nucleation due to any semi-classical instability of Kaluza-Klein vacuum as we observe in the Witten bubble case \cite{Witten:1981gj}. More specifically, we restrict our focus on the Kaluza-Klein compactification involved in M$5$-bubble solutions. The analyses for other cases will be straightforward.  In the end, we will see the bubbles {\it cannot} be considered as the remnants of unstable Kaluza-Klein vacua. Basically it is due to the fact that the bubbles are RR-charged, though imaginary, while Kaluza-Klein vacuum is neutral.

Kaluza-Klein compactification M$^{9+1}\times $S$^{1}$ is classically stable (against small oscillations). In order to check its semi-classical stability, that is, to check any possible penetration to other vacua, we resort to the analysis in Euclidean space. This method is summarized as two steps. First, look for a bounce solution (a Euclidean solution asymptotically approaching M$^{10}\times$ S$^{1}$). One way to find the Euclidean solution is to take Wick rotation on a known solution in M$^{10+1}$ and make the formerly temporal coordinate compact with appropriate periodicity to achieve the geodesic completeness. Second, check the possibility of the negative modes for small oscillations around the solution. Instead of this latter step, one can directly find the solution into which the `false vacuum' decay \cite{Witten:1981gj}. The solution turns out to be Lorentzian solution found by the analytic continuation from Euclidean bounce solution found in the first step. The upshot is that the solution into which the geometry M$^{9+1}\times $S$^{1}$ tunnels into semi-classically can be found by taking DWR on a known solution in M$^{10+1}$ dimensions. 

Since M2- or M5-bubble geometry are found in that way, there might be a possiblity of $M^{9+1}\times S^1$'s being semi-classically unstable decaying into M2- or M5-bubble geometry.
However, one very important thing in the argument is whether these new geometries have the same energy, that is vanishing, as that of the `would-be' unstable M$^{9+1}\times$ S$^{1}$ vacuum. They do not seem so. We first note that the relation between M-theory in eleven dimensions and the string theory in ten dimensions is not like ordinary Kaluza-Klein style. In order to make sense of ten dimensional theory by the compactification, we have to use a very specific compactification scheme. Regarding the direction $\xi$ as the eleventh dimension circle, the resulting geometry is D$4$-bubble in IIA;
\begin{eqnarray}\label{e47}
ds^{2}&=& e^{ \frac{\phi}{2}}\,ds^{2}_{E}=H^{- \frac{1}{2}}(r) \sum^{5}_{i=1}(dx^{i})^{2}+H^{ \frac{1}{2}}(r) \left( -r^{2}d\psi^{2}+dr^{2}+r^{2}\cosh^{2}\psi\,d\Omega^{2}_{3}\right)  \nonumber\\
e^{\phi(r)}&=&H^{- \frac{1}{4}}(r), \qquad F^{(4)}=-3i\,Q\cosh^{3}\psi \,d\psi\wedge d\Omega_{3}, \qquad H(r)=1+ \frac{Q}{r^{3}}.
\end{eqnarray}
The energy-momentum tensor (computed in ten dimensional Einstein frame metric $ds^{2}_{E}$) is
\begin{eqnarray}\label{e48}
T_{ \mu\nu}&=& \frac{1}{2} \left(\partial_{\mu}\phi\partial_{\nu}\phi- \frac{g_{E\, \mu\nu}}{2}\partial_{\lambda}\phi\partial^{\lambda}\phi \right)  \nonumber\\
&&+ \frac{e^{\frac{\phi}{2}}}{2\cdot 4!} \left(4 F_{\mu\rho_{1}\rho_{2}\rho_{3}}F_{\nu}{}^{\rho_{1}\rho_{2}\rho_{3}} - \frac{g_{E\, \mu\nu}}{2}F_{\rho_{1}\rho_{2}\rho_{3}\rho_{4}}F^{\rho_{1}\rho_{2}\rho_{3}\rho_{4}}\right). 
\end{eqnarray}
Computing the component $T_{\psi\psi}$, we get the following result;
\begin{eqnarray}\label{e49}
T_{\psi\psi}&=& \frac{r^{2}H^{-2}}{4\cdot 16}\left(\partial_{r}H \right)^{2}+ \frac{H^{-\frac{1}{8}}}{4} \left( H^{ \frac{5}{8}}r^{2}\cosh^{2}{\psi}\right)^{-3}\left(F_{\psi\vec{\Omega}} \right)^{2}   \nonumber\\
&=&\frac{9Q^{2}r^{-6}H^{-2}}{4\cdot 16}-\frac{9Q^{2}r^{-6}H^{-2}}{4}=-\frac{3^{3}\cdot 5 Q^{2}r^{-6}H^{-2}}{2^{6}}.
\end{eqnarray}
The contribution from the imaginary RR four form field reuslts in the negative energy in lower dimensions. 

The energy difference between the Kaluza-Klein vacuum and the bubble solutions implies that the latter is not due to any semi-classical instability of the former. This becomes indeed clear if we see RR-form flux that the bubbles carry. There is no such a conserved (though imaginary valued) RR-form flux involved in the Kaluza-Klein vacuum. Moreover, the negative energy of the solution read off in ten-dimensions is not actually concerned with the compactification. In the next subsection, we check the energy condition of M$5$-bubble solutions directly in eleven dimensions. 
 
\subsection{Energy Condition of the Bubbles}

The energy-momentum tensor can be easily read from the supergravity action
\begin{equation}\label{11sugra}
S= \frac{1}{16\pi G^{11}_{N}}\int d^{11}x \sqrt{-g} \left(R- \frac{1}{2\cdot 4!}|\mathcal{F}_{(4)}|^{2}- \frac{1}{6\cdot 4!\cdot4!\cdot3!}\epsilon\cdot \mathcal{F}_{(4)}\cdot \mathcal{F}_{(4)}\cdot C_{(3)} \right). 
\end{equation}  
 as
\begin{equation}\label{e50}
T_{ \mu\nu}= \frac{1}{2\cdot 3!}\mathcal{F}_{\mu\rho\sigma\kappa}\mathcal{F}_{\nu}{}^{\rho\sigma\kappa}- \frac{g_{ \mu\nu}}{2\cdot 2\cdot 4!}\mathcal{F}_{\lambda\rho\sigma\kappa}\mathcal{F}^{\lambda\rho\sigma\kappa}.
\end{equation}  
Since 
\begin{eqnarray}\label{e51}
\mathcal{F}_{\lambda\rho\sigma\kappa}\mathcal{F}^{\lambda\rho\sigma\kappa}
&=&4! g^{\psi\psi}g^{\vec{\Omega}_{3}\vec{\Omega}_{3}} \left( -3i Q \cosh^{3}{\psi}\right)^{2} \nonumber\\
&=&-4! H^{- \frac{2}{3}}r^{-2} \left( H^{-\frac{2}{3}}r^{-2}\cosh^{-2}\right)^{3}\left(-3i Q \cosh^{3}{\psi} \right)^{2} \nonumber\\  
&=&4!\cdot 9 Q^{2} \left( H^{-\frac{2}{3}}r^{-2}\right)^{4},
\end{eqnarray}
and 
\begin{eqnarray}\label{e52}
\mathcal{F}_{\mu\rho\sigma\kappa}\mathcal{F}_{\nu}{}^{\rho\sigma\kappa}
&=& \left\{\begin{array}{ll}0 & \qquad (\,\mu,\,\nu\in \{\xi,\,\vec{x},\,r\,\}\,)\\3!\cdot 9Q^{2}\left( H^{- \frac{2}{3}}r^{-2}\right)^{4}g_{ \mu\nu}  &\qquad (\,\mu,\,\nu\in \{\psi,\,\vec{\Omega}\}\,) \end{array}\right.
\end{eqnarray}
the components of the energy-momentum tensor will be
\begin{eqnarray}\label{energycondition}
T_{\mu\nu}&=&\left\{\begin{array}{cc} - \frac{9Q^{2}}{4}\left( H^{- \frac{2}{3}}r^{-2}\right)^{4}g_{ \mu\nu}& \qquad (\,\mu,\,\nu\in \{\xi,\,\vec{x},\,r\,\}\,) \\ \frac{9Q^{2}}{4}\left( H^{- \frac{2}{3}}r^{-2}\right)^{4}g_{ \mu\nu} & \qquad (\,\mu,\,\nu\in \{\psi,\,\vec{\Omega}\}\,) \end{array}\right.
\end{eqnarray}
The expression becomes more succinct in the orthonormal frame;
\begin{equation}\label{e53}
T_{ab}=\frac{9Q^{2}}{4}H^{- \frac{8}{3}}r^{-8}\text{diag} \left(\underbrace{-1}_{\xi},\,\underbrace{-1,\,-1,\,-1,\,-1,\,-1}_{\vec{x}},\,\underbrace{-1}_{r},\,\underbrace{-1}_{\psi},\,\underbrace{1,\,1,\,1}_{\vec{\Omega}}\right). 
\end{equation}   
The energy density $\rho$ is negative and the pressure $p$ along Euclidean world volume and the radial direction is negative while the pressure along 3-sphere is positive. The geometry therefore violates both the weak energy condition (because $T_{00}<0$) and the strong energy condition (because $T_{00}-\eta_{00}T/2<0$). The null energy condition is violated for the null direction concerning the coordinates $(\xi,\,\vec{x},\,r)$ but is safe for the null direction involving the spherical coordinates. 

This result is in contrast with the case of D/M-branes. The energy-momentum tensor of M$5$-branes in the orthonormal frame is
\begin{equation}\label{e54}
T_{ab}=\frac{9Q^{2}}{4}H^{- \frac{8}{3}}r^{-8}\text{diag} \left(\underbrace{1}_{t},\,\underbrace{-1,\,-1,\,-1,\,-1,\,-1}_{\vec{x}},\,\underbrace{-1}_{r},\,\underbrace{1,\,1,\,1,\,1}_{\vec{\Omega}}\right), 
\end{equation}   
thus satisfies the weak energy condition ($T_{00}>0$), but violates the strong energy condition (because $T_{00}-\eta_{00}T/2<0$). It also satisfies the null energy condition ($T_{++}\geq 0$).

The equation of state parameter $w=p/\rho$ reveals very peculiar feature of the solution. It is either $1$ or $-1$ depending on the directions. 
In the near-bubble regime, the factor in front of the energy-momentum tensor becomes constant, that is,
\begin{equation}\label{e55}
\frac{9Q^{2}}{4}H^{- \frac{8}{3}}r^{-8}\simeq \frac{9Q^{- \frac{2}{3}}}{4}
\end{equation}  
 so that the geometry becomes factorized into 7-hyperboloid and $(3+1)$-de Sitter;
 \begin{equation}\label{e56}
R_{ab}=T_{ab}- \frac{1}{9}\eta_{ab}T=\frac{3}{2}Q^{- \frac{2}{3}}\text{diag} \left(\underbrace{-1}_{\xi},\,\underbrace{-1,\,-1,\,-1,\,-1,\,-1}_{\vec{x}},\,\underbrace{-1}_{r},\,\underbrace{-2}_{\psi},\,\underbrace{2,\,2,\,2}_{\vec{\Omega}}\right)
\end{equation}  
This implies that the equation of state parameter $w=-1$, read from the directions ($\psi,\,\vec{\Omega}$) of (\ref{energycondition}) in the near bubble regime, looks to be concerned with the cosmological constant. However, the sign of Einstein tensor ($G_{ab}=9Q^{-\frac{2}{3}}\eta_{ab}/4$) for dS$_{3+1}$ part in the full eleven dimensional solution, is opposite to that for pure $(3+1)$-dimensional de Sitter space-time ($G_{ab}=-3Q^{- \frac{2}{3}} \eta_{ab}$). The same is true for the anti-de Sitter compactification of D-brane geometry. Therefore we are not quite sure of the meaning of the equation of state parameter $w=-1$ in these kinds of compactifications. At the least, it is different from that of phantoms ($w<-1$) and is consistent with the empirical data on the dominent contribution of the matter component \cite{Perlmutter:1998np} \cite{Riess:1998cb}. Even though we have imaginary fields, thus can be regarded as fields with the kinetic term with the wrong sign, its trivial dynamics ($\mathcal{F}_{(4)}$ is constant) makes the equation of state parameter $w$ different from that of the ordinary phantom fields, and its status phenominologically safe in the sense discussed in Ref. \cite{Carroll:2003st}. 
 
\section{Extending Hartle-Hawking to Extra Dimensions}

We rather interpret the bubble solution just literally as the solution coming from the geometry, `anti-de Sitter $\times$ sphere' via DWR. In other word, we regard DWR not just as a mathematical tool (to generate a new solution from a given solution) but as a physical process like quantum tunneling. This is very analogous to Hartle-Hawking proposal on the creation of the Universe, which describes de Sitter space-time tunneling from the sphere \cite{HartleHawking}. The bubble solution adds the hyperbolic space as another ingredient. It tunnels from anti-de Sitter space-time. 

\subsection{Junction Condition for the Real Tunneling Solutions}
In order to be more precise about the surgery of Euclidean geometry and Lorentzian geometry, let us describe the situation in the language of the complex geometry. From here on, we will closely follow the argument developed in Ref. \cite{Gibbons:1990ns} and apply it to the case at hand. The real tunneling solutions, describing the transition from the purely Euclidean manifold to the purely Lorentzian manifold (and vice versa), require that the second fundamental form $K_{ij}$ vanish on the common boundary of those two manifolds because it is defined as $K_{ij}=\nabla_{[i}n_{j]}$ or $K_{ij}=\pm i \nabla_{[i}n_{j]}$, on each manifold, where $n^{\alpha}$ is the unit normal to the common boundary. One simple way to achieve this goal is to consider a complex geometry so that two involutive isometries determine two real sub-geometries (one Euclidean and the other Lorentzian) as their fixed point sets. The same involution maps can be used to define the common boundary of those two real sub-geometries. We give the details  below.   

The near-bubble geometry $2M_{I}\equiv $ dS$_{3+1}\times$ H$^{7}$ can be considered as the double manifold (obtained by joining two copies, $M^{+}_{I}$ and $M^{-}_{I}$ of a manifold $M_{I}$, across their common boundary $\Sigma_{I}$) defined in Ref. \cite{Gibbons:1990ns}. The near-horizon geometry S$^{4}\times$AdS$_{6+1}$ of the M$5$-brane cousin is another double manifold $2M_{II}$ appearing in our story. 

Both double manifolds are the fixed point sets of two anti-holomorphic involutions, respectively. Let us consider a complex manifold $\mathcal{M}_{5}\times \mathcal{M}_{8}$ and embed two hypersurfaces by the following two complex quadrics;
\begin{equation}\label{e57}
\sum^{5}_{i=1} Z^{2}_{i}=l^{2}_{4},\qquad \sum^{8}_{j=1} W^{2}_{j}=-l^{2}_{7}.
\end{equation}   
The manifold $2M_{I}$ is the fixed point set of the anti-holomorphic involution,
\begin{eqnarray}\label{involution1}
J_{I}: \, &&(Z_{1},\,Z_{2},\,\cdots, Z_{5};\,W_{1},\,W_{2},\,\cdots, W_{8}) \nonumber\\
&&\rightarrow (\bar{Z_{1}},\,\bar{Z_{2}},\,\cdots, -\bar{Z_{5}};\,-\bar{W_{1}},\,\bar{W_{2}},\,\cdots, \bar{W_{8}}),
\end{eqnarray}
while the manifold $2M_{II}$ is the fixed point set of another anti-holomorphic involution,
\begin{eqnarray}\label{involution2}
J_{II}: \, &&(Z_{1},\,Z_{2},\,\cdots, Z_{5};\,W_{1},\,W_{2},\,\cdots, W_{8}) \nonumber\\
&&\rightarrow (\bar{Z_{1}},\,\bar{Z_{2}},\,\cdots, \bar{Z_{5}};\,\bar{-W_{1}},\,\bar{W_{2}},\,\cdots, -\bar{W_{8}}).
\end{eqnarray}  

Both double manifolds compose two real slices embedded into the complex manifold $\mathcal{M}_{5}\times \mathcal{M}_{8}$ (`real' in the sense that the metrics fulled back on them are real valued). The intersection of these fixed point sets is therefore given by
\begin{equation}\label{e58}
\sum^{4}_{i=1}Z^{2}_{i}=l^{2}_{4}, \qquad -W^{2}_{1}+\sum^{7}_{j=2}W^{2}_{j}=-l^{2}_{7}
\end{equation}  
with all $Z_{i}$ ($i=1, 2,\cdots 4$) and $W_{j}$ ($j=1, 2,\cdots 7$) real, thus
defining S$^{3}\times$H$^{6}$. The anti-holomorphic involution $J_{II}$ acting on $2M_{I}$ exchanges $M^{+}_{I}$ with $M^{-}_{I}$, while the involution $J_{I}$ swaps $M^{+}_{II}$ with $M^{-}_{II}$ when it acts on $2M_{II}$. These correspond to the orientation-flip or time-reverse operations. According to the argument in \cite{Gibbons:1990ns}, there is no distributional constribution to Ricci tensor $R_{ab}$ at the intersection, S$^{3}\times$H$^{6}$, that is, the fixed point set of both involutions $J_{I}$ and $J_{II}$. Therefore, we can make the junction of $M^{+}_{I}$ (a half of dS$_{3+1}\times$H$^{7}$) and $M^{-}_{II}$ (a half of S$^{4}\times$AdS$_{6+1}$) across their common boundary S$^{3}\times$H$^{6}$.  

\subsection{the Surgery of S$^{2}\times$AdS$_{1+1}$ and dS$_{1+1}\times$H$^{2}$ as an Example}

To be more specific, let us take an example of the surgery S$^{2}\times$AdS$_{1+1}$ and dS$_{1+1}\times$H$^{2}$. For the aim, it is convenient to use the global coordinates which parametrize the sphere  and anti-de Sitter, that is $2M_{II}$, as
\begin{eqnarray}\label{e59}
\left.\begin{array}{ll}Z_{1}=\sin{\theta}\cos{\varphi},\quad Z_{2}=\sin{\theta}\sin{\varphi},\quad Z_{3}=\cos{\theta}&\qquad(\text{S}^{2})\\
W_{1}=\cosh{\lambda}\cos{\tau},\quad W_{2}=\sinh{\lambda},\quad W_{3}=\cosh{\lambda}\sin{\tau}&\qquad(\text{AdS}_{1+1})
\end{array}\right.
\end{eqnarray}
so that the embedding coordinates satisfy $Z^{2}_{1}+Z^{2}_{2}+Z^{2}_{3}=1$ (sphere) and $-W^{2}_{1}+W^{2}_{2}-W^{2}_{3}=-1$ (anti-de Sitter). The coordinates for the hyperboloid and de Sitter, viz., $2M_{I}$ in the above, will be
\begin{eqnarray}\label{e60}
\left.\begin{array}{ll}Z_{1}=\cosh{\psi}\cos{\varphi},\quad Z_{2}=\cosh{\psi}\sin{\varphi},\quad Z_{3}=\sinh{\psi}&\qquad(\text{dS}_{1+1})\\
W_{1}=\cosh{\lambda}\cosh{\mu},\quad W_{2}=\sinh{\lambda},\quad W_{3}=\cosh{\lambda}\sinh{\mu}&\qquad(\text{H}^{2})\end{array}\right.
\end{eqnarray} 
These latter sets satisfy $Z^{2}_{1}+Z^{2}_{2}-Z^{2}_{3}=1$ (de Sitter) and $-W^{2}_{1}+W^{2}_{2}+W^{2}_{3}=-1$ (hyperboloid). The involutions $J_{I}$ and $J_{II}$ defined in (\ref{involution1}) and (\ref{involution2}) leave the doubles $2M_{I}$ and $2M_{II}$, invariant respectively. However, the involution $J_{I}$ acting on $2M_{II}$, and $J_{II}$ acting on $M_{I}$ result in the changes $Z_{3}\rightarrow -Z_{3}$ and $W_{3}\rightarrow -W_{3}$. The fixed points (where $Z_{3}=W_{3}=0$) of these mappings are specified by $\tau=-i\mu=0$ and $\theta=i\psi+\pi/2=\pi/2$, where one can make the junction of $M^{-}_{II}$ ($\tau<0$, $\theta>\pi/2$) and $M^{+}_{I}$ ($\psi>0$, $\mu>0$). 

After the surgery, the geometry looks like Fig. 8. The left figure (a) shows the anti-de Sitter tunneling to the hyperbolic space while the right fugure (b) illustrates the familiar Hartle-Hawking tunneling from the sphere to de Sitter. 

\FIGURE{\epsfig{file=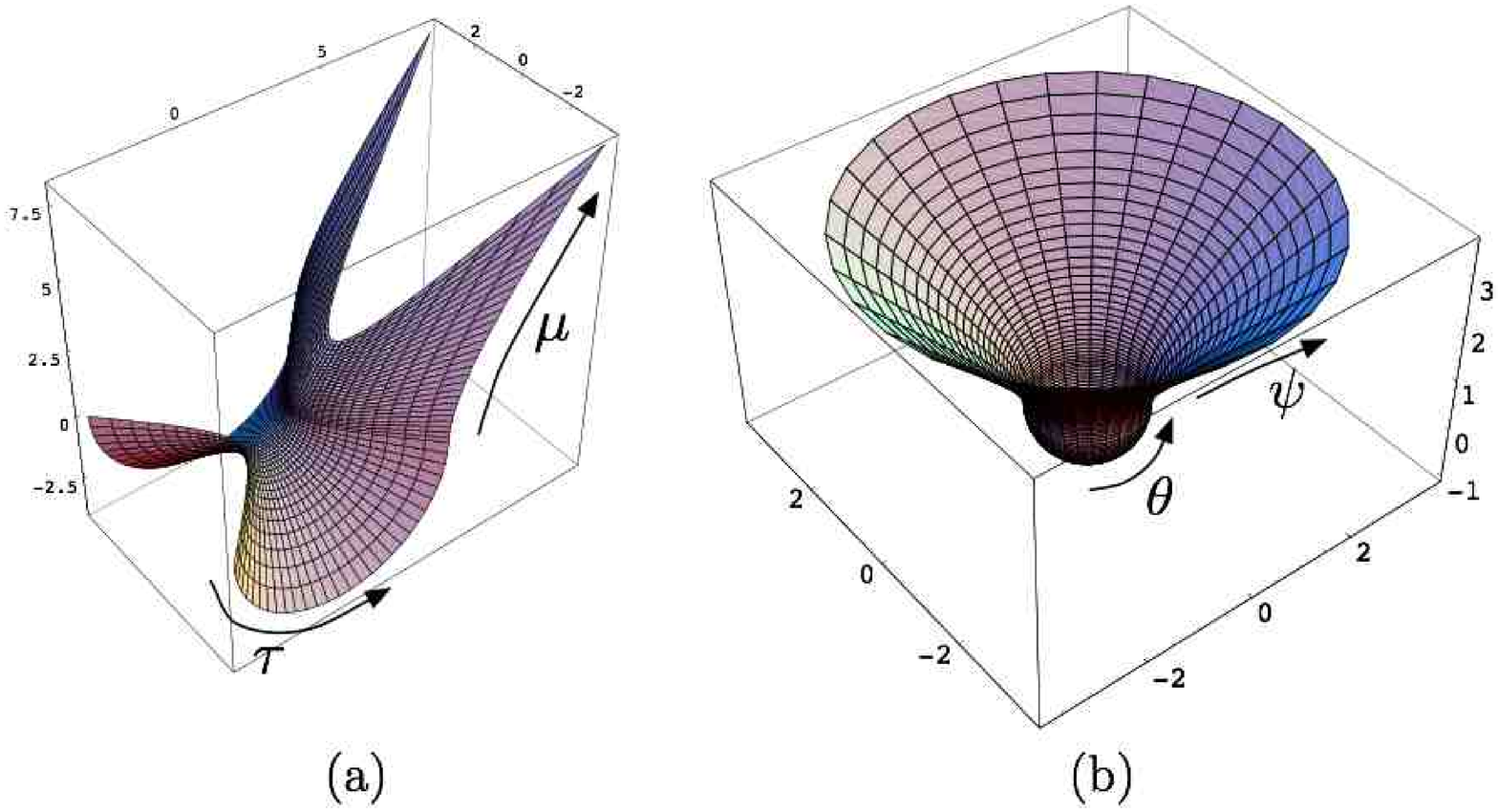,width=11cm} 
        \caption{Extension of Hartle-Hawking: (a) anti-de Sitter tunnels to a hyperbolic space,  (b) the familiar Hartle-Hawking tunneling process from a sphere to de Sitter}
	\label{figure8}}

%\begin{figure}[htbp]
%\begin{center}
%\includegraphics{hartlehawking.pdf}
%\caption{Extension of Hartle-Hawking: (a) anti-de Sitter tunnels to a hyperbolic space,  (b) the familiar Hartle-Hawking tunneling process from a sphere to de Sitter}
%\label{figure8}
%\end{center}
%\end{figure}

There are some features to note in this example. First, the actual anti-de Sitter space-time is the universal covering space of the corresponding part shown in the figure. Second, the coordinates $(\lambda,\,\mu)$ in the hyperbolic part are different from the coordinates $(r,\,\xi)$ used so far. The part glued to anti-de Sitter space-time in the figure is Poincar\'{e} (Lobachevsky) half-plane (H$^2/\mathbf{Z}_2$; where $\mathbf{Z}_2$ is $J_{II}$ dicussed above) and therefore the angular coordinate $\xi$ runs from $0$ to $\pi$, that is to say, Kaluza-Klein internal space is not a circle but an interval (S$^{1}/\mathbf{Z}_{2}$). In other words, the `AdS Universe' is destined to the hyperboloid compactified on a half circle S$^{1}/\mathbf{Z}_{2}$. This feature is common to all other bubble solutions discussed in this paper. Especially the standard M-compactification of the M$2$- or M$5$-bubble backgrounds will lead to the heterotic bubble vacuum in $10$-dimensions. 
Third, we have to worry about the radial fluctuation (along $r$) which could invalidate the compactification scheme. Fortunately, this can be controlled by considering the compact (closed) hyperbolic space. General $n$-dimensional hyperboloid and anti-de Sitter manifold have common isometry subgroup $SO(n-1, 1)$ over their common boundary H$^{n-1}$. In order to stabilize the radial fluctuation, we replace the previous common boundary H$^{n-1}$ with the compact hyperbolic space H$^{n-1}/\Gamma$, where $\Gamma=SO(n-1, 1;\mathbf{Z})$. Then all the radial modes will be confined to Lobachevsky space. See Refs. \cite{Kaloper:2000jb} \cite{Nasri:2002rx} for details about the compact hyperbolic extra dimensions and the radion stabilization.    

\subsection{the Wave Function of the Universe}

 At this point, some might be worried about the transition from anti-de Sitter to the hyperbolic space because it looks contradictory to the fact that anti-de Sitter spacetime is quite stable \cite{Breitenlohner:1982bm}. However, we have to note the precise meaning of the stability of anti-de Sitter; it is stable for fluctuations which vanish sufficiently fast at spatial infinity, and can be stable even when the scalar potential (the sort appearing in the gauged supergravity) is unbounded below. It is obvious that this fact does not exclude other possibilities of instability caused by thermal effect or some non-perturbative quantum tunneling like topology change. In fact, anti-de Sitter spacetime has Jeans instability \cite{Ginsparg:1982rs}. 
 
 In this paper, we are talking about the latter possibility; the non-perturbative quantum tunneling. In fact, the hyperbolic space, into which the anti-de Sitter spacetime is Wick-rotated, can be regarded as Euclidean geometry to define the ground (least excited) state wave function of the Universe with the negative cosmological constant \`{a} la Hartle-Hawking \cite{HartleHawking}. This can be made more explicit as follows. 
 
Let us first summarize briefly the result of no-boundary proposal\cite{HartleHawking} for the wave function of the Universe.
According to the proposal, the ground state ampitude of a $3$-geometry is Euclidean path integral over all compact positive definite $4$-geometries with the $3$-geometry as a boundary. 
Its extension to the general $(n+1)$-dimensional universe with a specific $n$-geometry $h_{ij}$ at a spatial section of the universe will be Euclidean path integral over all compact positive definite $(n+1)$-geometries $g_{ \mu\nu}$ with $h_{ij}$ as a boundary;
\begin{equation}\label{e61}
\Psi_{0}[h]\sim\int_{g}e^{-I_{E}(g)}.
\end{equation}    

Let us apply this idea to our case.
DWR can be understood from two different view points; anti-de Sitter and de Sitter. It can be summarized as the change of temporal coordinate from the AdS time, $\tau$, to the dS time, $\psi$ at $\tau=0$ and $\psi=0$. 
In view of AdS time, a quantum state of the geometry is defined on a spatial section H$^6\times(\mbox{S}^4/\mathbf{Z}_2)$ characterized by constant $\tau(\leq 0)$. In the half four-sphere, S$^4/\mathbf{Z}_2$, the latitudinal coordinate $\theta$ runs from $\pi/2$ to $\pi$. Its corresponding wave function will be 
\begin{eqnarray}
\Psi^*_{\mbox{AdS}}\left[h_{ij}\left(\mbox{H}^6\times\left(\mbox{S}^4/\mathbf{Z}_2\right)\right)\right]\sim\int_C\delta g\, e^{-I[g]},
\end{eqnarray}  
where the integration is over the set $C$ of all eleven-geometries on $(\mbox{H}^7/\mathbf{Z}_2)\times (\mbox{S}^4/\mathbf{Z}_2)$ but with the metric on the half four-sphere part kept fixed (to be that of a pure round four-sphere). Here complex conjugate, which affects only the overall normalization constant, was taken to account for the time reversal (the death of the `AdS Universe'). Euclidean action functional $I$ is obtained by Wick rotation, $\mu\equiv i\tau$, where the coordinate $\mu$ should be understood as Euclidean time rather than a spatial coordinate of the compact seven-hyperboloid. We assumed a proper compactification of the seven hyperboloid part to employ Hartle-Hawking no (final) boundary conjecture. 

In view of dS time, the spatial section on which a quantum state is defined, will be $(\mbox{H}^7/\mathbf{Z}_2)\times\mbox{S}^3$. The ground state wave function of de Sitter Universe is given by
\begin{eqnarray}
\Psi_{\mbox{dS}}\left[h_{ij}\left(\left(\mbox{H}^7/\mathbf{Z}_2\right)\times\mbox{S}^3\right)\right]\sim\int_{C'}\delta g\, e^{-I[g]},
\end{eqnarray}  
where $C'$ is the set of all eleven-geometries on $(\mbox{H}^7/\mathbf{Z}_2)\times (\mbox{S}^4/\mathbf{Z}_2)$ but with the metric on the seven-hyperboloid part kept fixed. Here, the latitudinal coordinate $\theta$ of four-sphere is no longer spatial coordinate but Euclidean time defined by $\theta=i\psi+\pi/2$. Meanwhile, the coordinate $\mu$, formerly known as the Euclidean time, is now understood as a spatial coordinate of the seven-hyperboloid.

It looks very difficult to imagine the transition from AdS$_{6+1}/\mathbf{Z}_2\times\mbox{S}^4/\mathbf{Z}_2$ to H$^7/\mathbf{Z}_2\times\mbox{dS}_{3+1}/\mathbf{Z}_2$ because the wave functions $\Psi_{\mbox{AdS}}$ and $\Psi_{\mbox{dS}}$ are those of ten-geometries defined on different topologies; the former is on $\mbox{H}^6\times\mbox{S}^4/\mathbf{Z}_2$ while the latter is on $\mbox{H}^7/\mathbf{Z}_2\times\mbox{S}^3$. However, one has to note that they are two different boundaries of the same topology. In fact,
\begin{eqnarray}
\partial\left(\left(\mbox{H}^7/\mathbf{Z}_2\right)\times\left(\mbox{S}^4/\mathbf{Z}_2\right)\right)=\mbox{H}^6\times\left(\mbox{S}^4/\mathbf{Z}_2\right)+\left(\mbox{H}^7/\mathbf{Z}_2\right)\times\mbox{S}^3.
\end{eqnarray} 
Therefore the transition can be described by
\begin{eqnarray}
<\Psi_{\mbox{dS}}\left[h_{ij}\left(\left(\mbox{H}^7/\mathbf{Z}_2\right)\times\mbox{S}^3\right)\right]\,\vert\,\Psi^*_{\mbox{AdS}}\left[h_{ij}\left(\mbox{H}^6\times\left(\mbox{S}^4/\mathbf{Z}_2\right)\right)\right]>\sim\int_\mathcal{C}\delta g\, e^{-I[g]},\quad
\end{eqnarray}   
where $\mathcal{C}$ is the set of all ten-geometries on $(\mbox{H}^7/\mathbf{Z}_2)\times(\mbox{S}^4/\mathbf{Z}_2)$.

Lastly, let us comment on the necessity of compactifying the hyperboloid. 
The main problem with the no-boundary proposal about the wave function of anti-de Sitter Universe is that its Euclidean partner (the hyperboloid) is not closed manifold. (As we discussed earlier, the same reason destabilizes the radion in the hyperbolic extra-dimensions.) In order to cure this problem, we have to make the hyperboloid compact, as before.
Without this procedure, we would have to worry about the boundary condition on the asymptotic boundary of the hyperboloid. Another reason for the compactification of the hyperboloid is to make lower dimensional Newton constant nonvanishing. There is a technical reason for this too. One can naively use the Euclidean solution characterizing $(\mbox{H}^7/\mathbf{Z}_2)\times(\mbox{S}^4/\mathbf{Z}_2)$ to compute the action $I(g)$. However, this scheme will fail because the result will be divergent, thereby gives null result for the transition we are thinking of. In fact, the whole Euclidean action with the solution plugged in is 
\begin{eqnarray}\label{e62}
I&=& - \frac{1}{16\pi G_{N}}\int d^{11}x_{E} \sqrt{g_{E}}\left( -R+ \frac{1}{2\cdot 4!}|\mathcal{F}_{(4)}|^{2}\right) \nonumber\\
&=&\frac{2}{4!\cdot 3!\cdot 16\pi G_{N}}\int d^{11}x_{E}\sqrt{g_{E}}\,|\mathcal{F}_{(4)}|^{2}\nonumber\\
&=&\frac{2\cdot 9\cdot 2\pi^{2} }{4\cdot 3!\cdot 16\pi G_{N}}\int d\xi \,d\vec{x}\, dr\, r^{2}\sim \infty.
\end{eqnarray} 

At least in (2+1)-dimensions, thus compactified hyperboloid becomes Riemann surface with genus $g\geq 2$ and gives a finite answer \cite{Fujiwara:1991tn} \cite{Oliveira-Neto:1997ap}. We will elaborate on this for higher dimensional cases in the forthcoming paper \cite{cho3}.

\section{Discussions}\label{seciii}

In this paper, we showed that most well-known D/M-brane configurations have their bubble cousins.
They have a universal feature of having de Sitter space-times as near-bubble geometries.
Especially in the extremal cases, these near-bubble de Sitter space-times are exact solutions of 10- or 11-dimensional supergravity theories.

The bubble solutions are obtained by taking DWR on D-branes or M-branes. Especially D-bubbles can also be obtained via timelike T-duality. For example, the extremal D3-bubbles are nothing but Hull's E4-branes in IIB theory. As for M-bubbles, DWR is a powerful tool to generate bubble solutions because there is no notion of T-duality in M-theory.

It is surprising that the near-bubble geometries of the extremal M$5$-, M$2$-, or D$3$-bubbles preserve 32 supersymmetries (for D$1$-D$5$-bubble case, $16$ supersymmetries). The imaginary higher-form fields guarantee Killing spinors even in de Sitter backgrounds. Therefore, the bubble solution is the domain wall solution interpolating two maximally supersymmetric regions, i.e., the asymptotic flat Minkowski region and  the near-bubble dS $\times$ H region. 

Especially supersymmetry makes sense of multi-bubble solutions obtained from multi-brane solutions via DWR. Though the observer can see the bubbles start inflating, once they invade the region near to the observer, each bubble will remain just as a deep throat of de Sitter space-time and it will take infinite coordinate time (affine parameter) for the bubbles to swallow up the whole space-time. At the final stage, the observer will find himself in a grand Swiss-cheese universe, where the effective cosmological constant varies from point to point near the cheese holes. This is a stringy realization of the multiverse idea that has drawn much interests \cite{Everett:1957hd} \cite{DeWitt:1973} \cite{Deutsch:1997}. The inflating bubbles leave behind the landscape of de Sitter skeleton. 

A fine virtue of this Swiss-cheese uiniverse is that there is a mechanism to control the cosmological constant to almost vanishing values.  
The imaginary field strengths triggers the creation of spherical D$3$-, M$5$-, M$2$ branes over the spatial section (S$^{4}$, S$^{6}$, S$^{3}$) of de Sitter space-time near D$3$-, M$2$-, M$5$-bubbles, respectively. As thus-created spherical branes condense, their tensions decelerate the expansion of de Sitter lowering the cosmological `constant' effectively. Though the effective cosmological constant decreases with the condensation of the spherical branes, the conservation of the fluxes `$\pm i$' of higher-form fields obstructs it to be completely vanishing. 

Lastly, we give a few comments on some topics of keen interests. 
\begin{itemize}
\item{Space-time Dimensionality}\\
The de Sitter compactification near bubble walls provide us with a natural explanation about why our space-time is of four dimensions. The M$5$-bubble solution is singled out of other bubble solutions as a model describing our Universe. It actually has some distinctive features from others. Together with the M$2$-bubble, it is a solution of $11$-dimensional supergravity, the low energy limit of M-theory that is currently considered as `the theory of everything'. Although both M-bubbles are geodesically complete solutions, the M$5$-bubble is more peculiar in the sense that it comes from the only non-singular solitonic solution, that is, the M$5$-brane. By `solitonic' we mean, the M$5$-brane solution does not require any source term in the action. Meanwhile, the M$2$-brane solution has a degenerate horizon inside which there is a sigularity. See   \cite{Duff:1994fg} for details. Though this inside region is excluded nicely in the M$2$-bubble solution, we would rather consider the M$5$-bubble as the solution describing the birth of our Universe from `nothing' (without any source). In this sense, $4$-dimensional space-time is more natural.   

\item{dS/CFT Duality}\\
Being different from other trials for de Sitter space-time in string/M-theories, the bubble geometry could provide a new setup for dS/CFT holography. Actually, the near-bubble geometry dS $\times$ H has three boundaries, one at the spatial infinity of the (non-compact) hyperbolic space and the others at two temporal boundaries of de Sitter space-time. In priciple, the holographic CFT of the bulk geometry can reside at any boundary of those three. However, since the original configuration before DWR had CFT at the boundary of AdS space-time and the information about the spherical part was captured there via global charges, it is likely that de Sitter part coming from the sphere will be in some way dual to the CFT on the boundary of the hyperbolic space. Such a dS/CFT would be different from what has been considered so far \cite{Strominger:2001pn}.

\end{itemize} 

\acknowledgments
We thank Seungjoon Hyun, Wontae Kim, Sunggeun Lee, and Boris Pioline for valuable comments and discussions.
This work is supported in part by Korea Research Foundation through Project No. KRF-2003-070-C00011.  It is also supported by the Science Research Center Program of the Korea Science and Engineering Foundation through 
the Center for Quantum space-time(CQUeST) of 
Sogang University with grant number R11-2005-021.
\begin{appendix}
\section{Action under the Double Wick Rotation (DWR)}
Let us investigate the generic behavior of the supergravity action under a certain multiple Wick rotation. We first note that any multiple Wick rotation can be described in the following compact form;
\begin{equation}\label{e63}
\tilde{e}^{a}=\lambda^{(a)}\,e^{a},\quad \tilde{\eta}_{ab}= \left(\lambda^{(a)} \right)^{-2}\eta_{ab}. 
\end{equation} 
Here the index $a$ is not dummy and $\lambda^{(a)}=\pm i$ depending on the scheme of Wick rotation. The result we derive below is valid even for the general scaling that needs not to be isotropic.    
The structure coefficient $C^{a}{}_{bc}$ defined by
\begin{equation}\label{e64}
de^{a}= \frac{1}{2}C^{a}{}_{bc}\,e^{b}\wedge e^{c}
\end{equation}  
transforms as
\begin{equation}\label{e65}
\tilde{C}_{abc}= \frac{1}{\lambda^{(a)}\lambda^{(b)}\lambda^{(c)}}\,C_{abc},
\end{equation}  
which leads to the transformation of the spin connection components;
\begin{equation}\label{e66}
\tilde{\omega}_{abc}= \frac{1}{2} \left(\tilde{C}_{abc}-\tilde{C}_{bac}-\tilde{C}_{cab} \right)=\frac{1}{\lambda^{(a)}\lambda^{(b)}\lambda^{(c)}}\,\omega_{abc}.
\end{equation}  
From the spin connection one form
\begin{equation}\label{e67}
\tilde{\omega}^{(1)}_{ab}=\tilde{\omega}_{abc}\tilde{e}^{c},
\end{equation}  
we note that
\begin{equation}\label{e68}
\tilde{\omega}^{(1)a}{}_{b}=\tilde{\eta}^{ac}\tilde{\omega}^{(1)}_{cb}= \left(\lambda^{(a)} \right)^{2} \eta^{ac} \frac{1}{\lambda^{(c)}\lambda^{(b)}}\omega^{(1)}_{cb}= \frac{\lambda^{(a)}}{\lambda^{(b)}}\omega^{(1)a}{}_{b},
\end{equation}  
 and Riemann tensor two form becomes
 \begin{eqnarray}\label{e69}
\tilde{R}^{(2)}_{ab}&=&d\tilde{\omega}^{(1)}_{ab}+\tilde{\omega}^{(1)}_{ac}\wedge \tilde{\omega}^{(1)c}{}_{b} =\frac{1}{\lambda^{(a)}\lambda^{(b)}}R^{(2)}_{ab},
\end{eqnarray}
 whose components transform as 
 \begin{equation}\label{e70}
\tilde{R}_{abcd}=\frac{1}{\lambda^{(a)}\lambda^{(b)}\lambda^{(c)}\lambda^{(d)}}R_{abcd}.
 \end{equation} 
 Ricci tensor transforms as
 \begin{equation}\label{e71}
\tilde{R}_{ac}= \frac{1}{\lambda^{(a)}\lambda^{(c)}}R_{ac}.
\end{equation}   
 Hence the curvature scalar is invariant under any Wick rotation;
 \begin{equation}\label{e72}
\tilde{R}=\tilde{\eta}^{ac}\tilde{R}_{ac}= \left(\lambda^{(a)} \right)^{2}\frac{1}{\lambda^{(a)}\lambda^{(c)}}\eta^{ac}R_{ac}=R\,. 
\end{equation}  
 We see this invariance of the curvature scalar under Wick rotations in the example of anti-de Sitter space-time and the hyperbolic space, also in the case of de Sitter space-time and the sphere. 
 
 Maxwell term is also invariant under any Wick rotation. Since
 \begin{eqnarray}\label{e73}
\tilde{F}^{(2)}&=& \frac{1}{2}\tilde{F}_{ab}\tilde{e}^{a}\wedge \tilde{e}^{b}= \frac{1}{2}\tilde{F}_{ab}\lambda^{(a)}\lambda^{(b)}e^{a}\wedge e^{b} \nonumber\\
&&\qquad\Rightarrow F_{ab}=\lambda^{(a)}\lambda^{(b)}\tilde{F}_{ab},
\end{eqnarray}
 one can then easily see 
 \begin{equation}\label{e74}
F_{ab}F^{ab}=\eta^{ac}\eta^{bd}F_{ab}F_{cd}= \left(\lambda^{(a)}\lambda^{(b)} \right)^{-2}\tilde{\eta}^{ac}\tilde{\eta}^{bd}\lambda^{(a)}\lambda^{(b)}\lambda^{(c)}\lambda^{(d)} \tilde{F}_{ab}\tilde{F}_{cd}=\tilde{F}_{ab}\tilde{F}^{ab}.
 \end{equation}  
 The same procedure can be applied to the higher form field to see that
 \begin{equation}\label{e75}
\tilde{F}^{(n)}_{a_{1}a_{2}\cdots a_{n}}= \frac{1}{\lambda^{(a_{1})}\lambda^{(a_{2})}\cdots\lambda^{(a_{n})}}\,F^{(n)}_{a_{1}a_{2}\cdots a_{n}}.
 \end{equation}  

Especially for the double Wick rotation, Jacobian factor in the volume is invariant because
\begin{equation}\label{e76}
\tilde{e}^{1}\wedge \tilde{e}^{2}\wedge\cdots \wedge\tilde{e}^{d}=\lambda^{(1)}\lambda^{(2)}\cdots\lambda^{(d)}e^{1}\wedge e^{2}\wedge \cdots\wedge e^{d}=e^{1}\wedge e^{2}\wedge \cdots \wedge e^{d}.
\end{equation}    
In the last step of the above equation, we used DWR prescription of $\tilde{e}^{0}=-i e^{0}$ and $\tilde{e}^{n}=i e^{n}$ so that $\lambda^{(1)}\lambda^{(2)}\cdots \lambda^{(d)}=1$. Actually, the same result is valid for arbitrary even number of Wick rotations as far as $\lambda^{(1)}\lambda^{(2)}\cdots \lambda^{(d)}=1$. Consequently, the action is invariant under arbitrary scaling or Wick rotation if $\lambda^{(1)}\lambda^{(2)}\cdots \lambda^{(d)}=1$. This symmetry of the action can be used as a method of generating a new solution from a given solution.

\section{D/M-bubbles from D/M-brane configurations via DWR}
{\bf B.1 D1-D5 Bubble}\

%\subsection{D1-D5 Bubble}

\noindent The non-extremal D1-D5 solution in $(5+1)$-dimensions is 
\begin{eqnarray}\label{e77}
ds^{2}_{E,6}&=&H^{- \frac{1}{2}}_{1}  H^{- \frac{1}{2}}_{5}\left(- f_{15}dt^{2} +dx^{2}\right) +H^{\frac{1}{2}}_{1}H^{\frac{1}{2}}_{5} \left(f^{-1}_{15K}dr^{2}+r^{2}d\Omega^{2}_{3}\right), \nonumber\\
A_{(2)}&=& \left(H'^{-1}_{1}-1 \right) dt\wedge dx,\qquad\mathcal{F}_{(3)}=-\partial_{r}H'_{5}\,d\Omega_{3}, \qquad e^{\phi_{6}}=1,
\end{eqnarray}
where
\begin{eqnarray}\label{e78}
&&H_{i}= 1+ \frac{\mu^{2}\sinh^{2}{\alpha_{i}}}{r^{2}}\quad (i=1,\,5),\qquad
H'_{5}= 1+ \frac{\mu^{2}\sinh{\alpha_{5}}\cosh{\alpha_{5}}}{r^{2}}, \nonumber\\
&&H'^{-1}_{1}-1=- \frac{\mu^{2} \sinh{\alpha_{1}}\cosh{\alpha_{1}}}{r^{2}+\mu^{2}\sinh^{2}{\alpha_{1}}},\qquad
f_{15}= 1- \frac{\mu^{2}}{r^{2}}.
\end{eqnarray}
The primed function is not to be confused with the derivative of the unprimed one. 

We take DWR used in (\ref{dwr}),
to get the following bubble solution,
\begin{eqnarray}\label{e79}
ds^{2}&=&H^{- \frac{1}{2}}_{1}  H^{- \frac{1}{2}}_{5}\left(f_{15}d\xi^{2}+dx^{2} \right) +H^{\frac{1}{2}}_{1}H^{\frac{1}{2}}_{5} \left(f^{-1}_{15}dr^{2}-r^{2}d\psi^{2}+r^{2}\cosh^{2}{\psi}d\Omega^{2}_{2}\right), \nonumber\\
A_{(2)}&=& -i\left(H'^{-1}_{1}-1 \right) d\xi\wedge dx,\qquad\mathcal{F}_{(3)}=-i\partial_{r}H'_{5}\,r^{3}\cosh^{2}{\psi}\,d\psi\wedge d\Omega_{2}. 
\end{eqnarray}

Taking the limit $u^{2}\ll \mu^{2}$ on the near-bubble coordinate $u^{2}\equiv r^{2}-\mu^{2}$, we get the near bubble geometry
\begin{eqnarray}\label{e80}
ds^{2}&\simeq& \left(\cosh{\alpha_{1}\cosh{\alpha_{5}}} \right)\left( du^{2}+ \frac{u^{2}}{\mu^{2}\cosh^{2}{\alpha_{1}}\cosh^{2}{\alpha_{5}}}d\xi^{2}+ \frac{1}{\cosh^{2}{\alpha_{1}}\cosh^{2}{\alpha_{5}}}dx^{2}\right)\nonumber\\
&&+\mu^{2}\cosh{\alpha_{1}}\cosh{\alpha_{5}}\left(-d\psi^{2}+\cosh^{2}{\psi}\,d\Omega^{2}_{2} \right)   \nonumber\\
\mathcal{F}_{(3)}&\simeq&-\frac{2i\sinh{\alpha_{1}}}{\mu^{2}\cosh^{3}{\alpha_{1}}}u\,du\wedge\,d\xi\wedge\,dx +2i\mu^{2}\sinh{\alpha_{5}}\cosh{\alpha_{5}}\cosh^{2}{\psi}\,d\psi\wedge d\Omega_{2}.
\end{eqnarray}

The extremal case is attained by limiting $\mu\rightarrow 0$ and $\alpha_{i}\rightarrow\infty$ so that $\mu^{2}\sinh{\alpha_{i}}\cosh{\alpha_{i}}$ $\simeq$ $\mu^{2}\sinh^{2}{\alpha_{i}}$ $\equiv Q_{i}$ be finite. Near the bubble boundary ($r\rightarrow 0$), the configuration becomes
\begin{eqnarray}\label{e81}
ds^{2}&=& \frac{r^{2}}{\sqrt{Q_{1}Q_{5}}} \left(d\xi^{2}+dx^{2} \right)+ \frac{\sqrt{Q_{1}Q_{5}}}{r^{2}}\left(dr^{2}-r^{2}d\psi^{2}+r^{2}\cosh^{2}{\psi}\,d\Omega^{2}_{2} \right)   \nonumber\\
\mathcal{F}_{(3)}&=&- \frac{2i\,r}{Q_{1}}dr\wedge d\xi\wedge dx+2iQ_{5}\cosh^{2}{\psi}\,d\psi\wedge d\Omega_{2}.
\end{eqnarray}

The electric and magnetic flux over the $(\xi,\,x,\,r)$-direction are respectively
\begin{eqnarray}\label{e82}
\int*\mathcal{F}_{(3)}&=&\int\frac{2iQ_{5}r }{\left(r^{2}+Q_{1} \right) \left(r^{2}+Q_{5} \right) }dr\wedge d\xi \wedge dx=\frac{i\ln{\frac{Q_{1}}{Q_{5}}}  }{\frac{Q_{1}}{Q_{5}}-1}\int d\xi\wedge dx, \nonumber\\
\int\mathcal{F}_{(3)}&=&-\int\frac{2iQ_{1}r}{\left(r^{2}+Q_{1} \right)^{2} } dr\wedge d\xi \wedge dx=-i\int d\xi\wedge dx.
\end{eqnarray}
The dependence of the electric flux on $Q_{1}$, as well as $Q_{5}$, is surprising. This is in contrast with D$1$-D$5$ branes on $T^{4}$, where the electric and the magnetic flux are given by
\begin{eqnarray}\label{e83}
\int_{S^{3}} *H_{(3)}&=&-2Q_{1}\lim_{r\rightarrow\infty}\int \frac{\left(r^{2}+Q_{1} \right) \left(r^{2}+Q_{5} \right) }{\left(r^{2}+Q_{1} \right)^{2} } d\Omega_{3}=-2Q_{1}\omega_{3},\nonumber\\
\int_{S^{3}} H_{(3)}&=&2Q_{5}\lim_{r \rightarrow \infty}\int d\Omega_{3}=2Q_{5}\omega_{3}.
\end{eqnarray}
Here $\omega_{n}$ stands for the volume of a unit $n$-sphere. Although the electric flux involves both $Q_{1}$ and $Q_{5}$, the final result depends only on $Q_{1}$.

%\subsection{M2-Bubble}

\vskip 0.2cm
\noindent
{\bf B.2 M2-Bubble}\\
\noindent
Non-extremal M2 solution takes the form;
\begin{equation}\label{e84}
ds^{2}=H_{T}^{- \frac{2}{3}}(r) \left(-dt^{2}f_{T}(r)+\sum^{2}_{i=1} dx^{2}_{i}\right)+H_{T}^{ \frac{1}{3}}(r) \left(dr^{2}f_{T}^{-1}(r)+r^{2} d\Omega^{2}_{7} \right),  
\end{equation}  
and $3$-form connection field is given by
\begin{equation}\label{e85}
C_{t12} =H_{T}'^{-1}-1=- \frac{\mu_{T}^{6}\sinh\alpha_{T}\cosh\alpha_{T}}{r^{6}}H_{T}^{-1}\,.
\end{equation}  
The explicit form of the harmonic functions are
\begin{eqnarray}\label{e86}
f_{T}(r)&=&1- \frac{\mu_{T}^{6}}{r^{6}},\quad
H_{T}(r)=1+ \frac{\mu_{T}^{6}\sinh^{2}\alpha_{T}}{r^{6}}.
\end{eqnarray}
$4$-form field strength constructed from the above connection is
\begin{equation}\label{e87}
\mathcal{F}_{(4)}=dC_{3}
= -\frac{6H_{T}^{-2}\mu_{T}^{6}\sinh\alpha_{T}\cosh\alpha_{T}}{r^{7}} dt\wedge dx^{1}\wedge dx^{2} \wedge dr.
\end{equation}

After double Wick rotations (one along the temporal direction and the other along one of the spherical directions) as in (\ref{dwr}),
we obtain the following bubble solution;
\begin{eqnarray}\label{e88}
ds^{2}&=&H_{T}^{- \frac{2}{3}}(r) \left(f_{T}(r) d\xi^{2}+dx_{1}^{2}+dx_{2}^{2} \right)+H_{T}^{ \frac{1}{3}}(r) \left(f_{T}^{-1}(r)dr^{2}-r^{2}d\psi^{2}+r^{2}\cosh^{2}\psi \,d\Omega_{6}^{2}\right) ,  \nonumber\\
\mathcal{F}_{(4)}&=& \frac{6i\,H_{T}^{-2}\mu_{T}^{6}\sinh\alpha_{T}\cosh\alpha_{T}}{r^{7}} d\xi\wedge dx^{1}\wedge dx^{2} \wedge dr\,.
\end{eqnarray}

In the near-bubble coordinate ($u^{2}=r^{6}-\mu_{T}^{6},\quad u^{2}\ll \mu_{T}^{6},$),
one can approximate the solution as
\begin{eqnarray}\label{e89}
ds^{2}&\simeq&\frac{\cosh^{ \frac{2}{3}}\alpha_{T} }{9\mu_{T}^{4}}\left(du^{2}+ \frac{9u^{2}}{\mu_{T}^{2}\cosh^{2}\alpha_{T} } d\xi^{2} \right)+\cosh^{- \frac{4}{3}}\alpha_{T} \left(dx_{1}^{2}+dx_{2}^{2} \right)\nonumber\\
&&\qquad+\mu_{T}^{2}\cosh^{ \frac{2}{3}}\alpha_{T} \left(-d\psi^{2}+\cosh^{2}\psi \,d\Omega_{6}^{2}\right) ,  \nonumber\\
\mathcal{F}_{(4)}&\simeq& \frac{2i\,\sinh\alpha_{T}}{\mu_{T}^{6}\cosh^{3}\alpha_{T}} u \,d\xi\wedge dx^{1}\wedge dx^{2} \wedge du\,.
\end{eqnarray}
In order for the geometry to be geodesically complete, the compact direction $(\xi\sim \xi+2\pi\hat{R})$ should have a period $\hat{R}=\mu_{T}\cosh\alpha_{T}/3$. For this value of $\hat{R}$, the geometry is factorized as $D_{2}\times \mathbf{ R}^{2}\times dS_{6+1} $, where the cosmological constant of de Sitter space-time is $\Lambda_{6+1}={15}/{\mu_{T}^{2}\cosh^{ \frac{2}{3}}\alpha_{T}}$.

Taking the limit $\mu_{T}\rightarrow 0$ and $\alpha_{T}\rightarrow \infty$ and keeping $Q_{T}=\mu_{T}^{6}\sinh\alpha_{T}\cosh\alpha_{T}$ finite, we obtain the extremal near-bubble configuration as an exact solution; 
\begin{eqnarray}\label{e90}
ds^{2}&=&\frac{Q_{T}^{ \frac{1}{3}} }{9\mu^{6}}\left(du^{2}+ \frac{9\mu_{T}^{4}u^{2}}{Q_{T} } d\xi^{2} \right)+ \frac{\mu_{T}^{4}}{Q_{T}^{ \frac{2}{3}}} \left(dx_{1}^{2}+dx_{2}^{2} \right)+Q_{T}^{ \frac{1}{3}} \left(-d\psi^{2}+\cosh^{2}\psi \,d\Omega_{6}^{2}\right) ,  \nonumber\\
\mathcal{F}_{(4)}&=& \frac{2i}{Q_{T}} u \,d\xi\wedge dx^{1}\wedge dx^{2} \wedge du\,.
\end{eqnarray}
In this case, the cosmological constant is $\Lambda=15/{Q^{ \frac{1}{3}}_{T}}$. 

The magnetic flux due to the M$2$-`instanton' that sources the extremal bubble geometry is
\begin{eqnarray}\label{e91}
\int_{ \text{H}_{4}}\mathcal{F}_{(4)}&=& \int\frac{6i\,Q_{T}} {r^{7}H^{2}_{T}} d\xi\wedge dx^{1}\wedge dx^{2} \wedge dr= \text{Vol}_{3}\int^{\infty}_{0} \frac{6i\,Q_{T} r^{5}\,dr}{\left( r^{6}+Q_{T}\right)^{2}} =i\, \text{Vol}_{3}.\nonumber
\end{eqnarray}

%\subsection{M5-Bubble}

\vskip 0.2cm
\noindent
{\bf B.3 M5-Bubble}\\
\noindent
Starting with the non-extremal M5 solution
\begin{equation}\label{e92}
ds^{2}=H_{F}(r)^{- \frac{1}{3}} \left(-dt^{2}f_{F}(r)+\sum^{5}_{i=1} dx^{2}_{i}\right)+H_{F}(r)^{ \frac{2}{3}} \left(dr^{2}f_{F}^{-1}(r)+r^{2} d\Omega^{2}_{4} \right),  
\end{equation}  
with the 4-form field
\begin{equation}\label{e93}
\mathcal{F}_{(4)}=*dH_{F}'=-3 \mu_{F}^{3}\sinh\alpha_{F}\cosh\alpha_{F}\,d\Omega_{4},
\end{equation}  
and taking DWR, we get 
\begin{eqnarray}\label{e94}
ds^{2}&=&\frac{H_{F}^{ \frac{2}{3}}(r)}{f_{F}(r)} \left(dr^{2}+\frac{f_{F}^{2}(r)}{H_{F}(r)} d\xi^{2}\right)+H_{F}^{- \frac{1}{3}}(r)\sum^{5}_{i=1}dx^{2}_{i} 
+H_{F}^{ \frac{2}{3}}(r) r^{2}\left(-d\psi^{2}+\cosh^{2}\psi\,\, d\Omega_{3}^{2}\right), \nonumber\\
\mathcal{F}_{(4)}&=&- 3i\mu_{F}^{3}\sinh\alpha_{F}\cosh\alpha_{F}\cosh^{3}\psi\,d\psi \wedge\,d\Omega_{3}.
\end{eqnarray}
In the above, the harmonic functions are
\begin{eqnarray}\label{e95}
f_{F}(r)=1- \frac{\mu_{F}^{3}}{r^{3}},\quad 
H_{F}(r)=1+ \frac{\mu_{F}^{3}\sinh^{2}\alpha_{F}}{r^{3}},\quad
H_{F}'(r)=1+ \frac{\mu_{F}^{3}\sinh\alpha_{F}\cosh\alpha_{F}}{r^{3}}.\nonumber
\end{eqnarray}

Introducing the near-bubble coordinate as $u^{2}\equiv r^{3}-\mu_{F}^{3}$, and considering the region of $u^{2}\ll \mu_{F}^{3}$, we get the geometry
\begin{eqnarray}\label{e96}
ds^{2}&\simeq& \frac{4\cosh^{ \frac{4}{3}}\alpha_{F}}{9\mu_{F}} \left(du^{2}+ \frac{9u^{2}}{4\mu_{F}^{2}\cosh^{2}\alpha_{F}} d\xi^{2} \right)+ \cosh^{- \frac{2}{3}}\alpha_{F}\sum^{5}_{i=1}dx_{i}^{2}  \nonumber\\
&&+\mu_{F}^{2}\cosh^{ \frac{4}{3}}\alpha_{F}\left(\cosh^{2}\psi\,\, d\Omega_{3}^{2} -d\psi^{2}\right). 
\end{eqnarray}
In order to make the geometry geodesically complete, we have to set the period of the angular coordinate as
$\xi\sim\xi+2\pi\hat{R}$, with $\hat{R}=(2\mu_{F}\cosh\alpha_{F})/3$. The near-bubble geometry is therefore $D_{2}\times \mathbf{R}^{5}\times dS_{3+1}$. 

Taking the extremal limit as $\mu_{F} \rightarrow0$ and $\alpha_{F} \rightarrow\infty$ with $Q\equiv\mu_{F}^{3}\sinh\alpha_{F}\cosh\alpha_{F}$ kept finite, and using a new radial coordinate,  $r=y^{2}$, we get 
\begin{eqnarray}\label{e97}
ds^{2}&=&4Q^{ \frac{2}{3}} \left( \frac{dy^{2}}{y^{2}}+ \frac{y^{2}}{4Q} \left(d\xi^{2}+\sum^{5}_{i=1} dx^{2}_{i}\right) \right)+Q^{ \frac{2}{3}} \left( -d\psi^{2}+\cosh^{2}{\psi}d\Omega^{2}_{3}\right), \nonumber\\
\mathcal{F}_{(4)}&=&-3iQ \cosh^{3}{\psi}\,d\psi \wedge\, d\Omega_{3}. 
\end{eqnarray}

The electric flux over the hyperbolic space is 
\begin{eqnarray}\label{e98}
\int_{ \text{H}_{7}}*\mathcal{F}_{(4)}&=&- 3i\,Q\int_{ \text{H}_{7}}\cosh^{3}\psi\,*\left(d\psi \wedge\,d\Omega_{3} \right) \nonumber\\
&=&-\int\frac{3i\,Q}{H^{2}_{F}r^{4}}d\xi\,\left(\wedge^{5}_{i=1} dx_{i}\right) \wedge dr \nonumber\\
&=&-3i\,Q\, \text{Vol}_{6}\int^{\infty}_{0} \frac{r^{2}}{\left(r^{3}+Q \right)^{2} }\,dr=-i\, \text{Vol}_{6}\,.
\end{eqnarray}

\section{Supersymmetry in the Near-Bubble Geometry: Details}

\vskip 0.2cm
\noindent
{\bf C.1 Global Coordinates for AdS$_{(n-1)+1}$ and H$^{n}$}\\
\noindent
Here, we introduce the global coordinates for anti-de Sitter space-time and the hyperbolic space. One important virtue of these coordinates is that they are dimensionless (a feature convenient for solving Killing spinor equations). 

Let us first consider AdS$_{3}$ case. Since it is a coset manifold $SO(2,\,2)/SO(2,\,1)$, one can pick up a hypersurface $-U^{2}-V^{2}+X^{2}+V^{2}=-R^{2}$ embedded in a flat space-time of signature $(2,\,2)$. We fix the length to be timelike so that the surface has $SO(2,\,1)$ isometry. (The little group of the timelike line element is $SO(2,\,1)$.) The global coordinates for this hypersurface are
\begin{eqnarray}\label{e99}
U&=&R\cosh{\lambda}\sin{\tau}, \nonumber\\
V&=&R\cosh{\lambda}\cos{\tau}, \nonumber\\
X&=&R\sinh{\lambda}\cos{\vartheta}, \nonumber\\
Y&=&R\sinh{\lambda}\sin{\vartheta}.
\end{eqnarray}
The flat metric induced on the hypersurface is
\begin{equation}\label{ads3m}
ds^{2}=R^{2} \left(-\cosh^{2}{\lambda}\,d\tau^{2}+d\lambda^{2}+\,\sinh^{2}{\lambda}\,d\vartheta^{2} \right). 
\end{equation}  

The hyperbolic space H$^{3}$, being the coset manifold $SO(3,\,1)/SO(3)$, is represented by the hypersurface
$\bar{U}^{2}-V^{2}+X^{2}+Y^{2}=-R^{2}$ embedded in a $(3+1)$-dimensional flat space-time. A proper parametrization for this hypersurface is obtained by Wick rotation $\tau=-i\mu$ from that of AdS$_{3}$ space-time;
\begin{eqnarray}\label{e100}
U&=&-i \,R\cosh{\lambda}\sinh{\mu}\equiv -i\bar{U}, \nonumber\\
V&=&R\cosh{\lambda}\cosh{\mu}, \nonumber\\
X&=&R\sinh{\lambda}\cos{\vartheta}, \nonumber\\
Y&=&R\sinh{\lambda}\sin{\vartheta}.
\end{eqnarray}
(Another Wick rotation $\tau=-i\mu +\pi/2$ is also possible, which amounts to exchange the role of $U$ and $V$.) From the metric $ds^{2}=d\bar{U}^{2}-dV^{2}+dX^{2}+dY^{2}$ of the ambient space-time, one can derive the metric induced on the hypersurface;
\begin{equation}\label{a1}
ds^{2}=R^{2} \left(\cosh^{2}{\lambda}\,d\mu^{2}+d\lambda^{2}+\,\sinh^{2}{\lambda}\,d\vartheta^{2} \right). 
\end{equation}

The above global coordinates (\ref{ads3m}) can be easily generalized to higher dimensional cases. The metric of AdS$_{(n-1)+1}$ space-time in the global coordinates is
\begin{equation}\label{a2}
ds^{2}=R^{2} \left(-\cosh^{2}{\lambda}\,d\tau^{2}+d\lambda^{2}+\,\sinh^{2}{\lambda}\,d\Omega^{2}_{n-2} \right). 
\end{equation}  
Wick rotation $\tau=-i\mu$ (or $\tau=-i\mu+\pi/2$) gives the metric of H$^{n}$.

\vskip 0.2cm
\noindent
{\bf C.2 Spin Connections for the Unit Sphere and the Unit de Sitter}\\
\noindent
In this subsection, we summarize the method to get spin connections for the symmetric spaces used in the paper. For more complete argument, see Ref. \cite{Ortin:2004ms}.
The $n$-sphere S$^{n}$ and the $n$-dimensional de Sitter space-time dS$_{(n-1)+1}$ are symmetric spaces diffeomorphic to the coset manifold $SO(n+1)/SO(n)$ and $SO(n,1)/SO(n-1,1)$ respectively.
Therefore every geometric information can be basically read off from the Lie algebras of those cosets;
\begin{eqnarray}\label{a3}
&&[P_{a},\,M_{bc}]=\eta_{ab}P_{c}-\eta_{ac}P_{b},\quad\qquad (a,\,b,\,c,\,d=1,\,2,\,\cdots , n)\\
&&[P_{a},\,P_{b}]=-M_{ab}\,,\quad [M_{ab},\,M_{cd}]=\eta_{bc}M_{ad}+\eta_{ad}M_{bc}-\eta_{ac}M_{bd}-\eta_{bd}M_{ac}\,.\nonumber
\end{eqnarray}  
Here $(\eta_{ab})=\text{diag}(1,\,1,\,\cdots,1,\,\zeta)$ with $\zeta=1$ for the sphere and $\zeta=-1$ for de Sitter. 

In these cases, Maurer-Cartan $1$-form constructed from the coset representative,
\begin{equation}\label{a4}
u(x^{1},\cdots,x^{n})=e^{x^{1}P_{1}}\cdots e^{x^{n}P_{n}},
\end{equation} 
is expanded as the sum of the vielbein $1$-form and the spin connection $1$-form. From now on in this section, whenever the products of the form $\prod^{q}_{a=p}f_{a}, \, (q<p)$ appears in any expression, we mean its value is equal to `$1$', just for the notational convenience. The explicit expression for Maurer-Cartan 1-form is
\begin{eqnarray}\label{left1form}
&&u^{-1}du=dx^{n}P_{n}+\sum^{n-1}_{a=1}dx^{a} P_{a} \left( \prod^{n-1}_{b=a+1}\cos{x^{b}}\right)\cos{\left(\sqrt{\zeta}x^{n} \right)} \nonumber\\
&&-\sum^{n-1}_{a=1}dx^{a}\left[M_{a\,n}\left( \prod^{n-1}_{b=a+1}\cos{x^{b}}\right) \frac{1}{\sqrt{\zeta}}\sin{\left( \sqrt{\zeta}x^{n}\right)}+\sum^{n-1}_{c=a+1}M_{a\,c}\left(\prod^{c-1}_{b=a+1}\cos{x^{b}} \right)\sin{x^{c}}  \right]\nonumber\\
&&\quad\qquad\!=\sum_{a=1}^{n}P_{a}e^{a}+\sum_{a<c}\,M_{ac}\,\omega^{ac}.
\end{eqnarray}  

The left invariant Euclidean metric on the coset space, being proportional to Killing metric, can be set to $B_{ab}=\eta_{ab}$. The geometry of the coset space, is then described by the following metric
\begin{eqnarray}\label{groupmetric}
ds^{2}=B_{ab}e^{a}e^{b}&=& \zeta\left(dx^{n} \right)^{2}+\left(dx^{n-1}\cos{\sqrt{\zeta}x^{n}}  \right)^{2}+\sum^{n-2}_{a=1} \left(dx^{a}\cos{\sqrt{\zeta}x^{n}}\prod^{n-1}_{b=a+1}\cos{x^{b}}  \right)^{2} \nonumber\\
&=& \zeta\, \left(e^{n}\right)^{2}+ \left(e^{n-1} \right)^{2}+\sum^{n-2}_{a=1} \left( e^{a}\right)^{2}.    
\end{eqnarray}
For the $n$-sphere ($\zeta=1$) or the $n$-dimensional de Sitter space-time ($\zeta=-1$), the above form of the metric can be related to the standard form by transforming the coordinates;
\begin{equation}\label{standardcoord}
x^{1}=\varphi,\qquad x^{a}=\theta_{a-1}- \frac{\pi}{2}\quad (a=2,\cdots n-1), \quad x^{n}=\left\{\begin{array}{cl}\theta_{n-1}-\pi/2 & (\zeta=1) \\\psi & (\zeta=-1)\end{array}\right.
\end{equation} 

\vskip 0.2cm
\noindent
{\bf C.3 Spin Connections in AdS$_{(n-1)+1}$ and H$^{n}$}\\
\noindent

The spin connections are rather easy to compute in the global coordinates if we make use of the results (\ref{left1form}) and (\ref{standardcoord}) for the unit $(n-2)$-sphere.  To be more specific, let us write down the vielbeins first;
\begin{eqnarray}\label{globalviel}
e^{0}&=& R\,\cosh{\lambda}\, dt, \nonumber\\
e^{1}&=& R\,d\lambda, \nonumber\\
e^{a}&=& R\,\sinh{\lambda}\,\tilde{e}^{a-1}\quad (2\le a \le n-1),
\end{eqnarray}
where $\tilde{e}^{a-1}$ denote the vielbeins of the unit sphere used in (\ref{groupmetric}) with the convention (\ref{standardcoord}). We note that the only difference between anti-de Sitter case and the hyperbolic case is the different signature of $\eta_{00}=\eta$ and the vielbeins take the same form in the above global coordinates with the change of $t \leftrightarrow \xi$ understood. Hereafter, we represent all the results collectively to be applicable for both cases. The frame structure equations $de^{a}=(C^{a}{}_{bc}\,e^{b}\wedge e^{c})/2$ are
\begin{eqnarray}\label{a5}
de^{0}&=& \frac{\tanh{\lambda}}{R}\,e^{1}\wedge e^{0}, \quad de^{1}=0, \nonumber\\
de^{a}&=& \frac{\coth{\lambda}}{R}\, e^{1}\wedge e^{a}+ \frac{1}{2R\sinh{\lambda}}\sum^{n-1}_{b,c\ge2}\tilde{C}^{a-1}{}_{b-1\,c-1}e^{b}\wedge e^{c},
\end{eqnarray}
where $\tilde{C}^{a-1}{}_{b-1\,c-1}$ denotes the frame structure constant of the unit $(n-2)$-sphere in the standard coordinates (\ref{groupmetric}) and the index $a$ runs from $2$ to $n-1$. Nontrivial frame structure constants are 
\begin{equation}\label{a6}
C^{0}{}_{10}= \frac{\tanh{\lambda}}{R},\qquad C^{a}{}_{1a}= \frac{\coth{\lambda}}{R},\qquad C^{a}{}_{bc}= \frac{1}{R\,\sinh{\lambda}}\tilde{C}^{a-1}{}_{b-1\,c-1}.
\end{equation}  
Using (\ref{globalviel}), we get the following non-vanishing  spin connection components;
\begin{eqnarray}\label{a7}
\omega_{01}&=& \frac{\eta}{R}\tanh{\lambda}\,e^{0}=\eta\,\sinh{\lambda}\,dt \nonumber\\
\omega_{1a}&=& - \frac{\coth{\lambda}}{R}\,e^{a}
%=-\cosh{\lambda}\,\tilde{e}^{a-1}
=- \cosh{\lambda}\, d\theta_{a-2}\prod^{n-3}_{b=a-1}\sin{\theta_{b}}, \quad(a=2,\cdots, n-2)\nonumber\\
\omega_{1\,n-1}&=& - \frac{\coth{\lambda}}{R}\,e^{n-1}
%=-\cosh{\lambda}\,\tilde{e}^{n-2}
=- \cosh{\lambda}\, d\theta_{n-3} \nonumber\\
\omega_{ab}&=& \frac{1}{R\sinh{\lambda}}\,\tilde{\omega}_{a-1\,b-1\,c-1}\,e^{c}
%=\tilde{\omega}_{a-1\,b-1\,c-1}\,\tilde{e}^{c-1}
=\tilde{\omega}_{a-1\,b-1},\quad
(a,\,b=2,\cdots,  n-1).
\end{eqnarray}
We denote $\varphi=\theta_{0}$ for notational convenience. 

The spin connections $\tilde{\omega}_{ab}\,\,(a<b)$ of the unit $(n-2)$-sphere are 
\begin{eqnarray}\label{a8}
\tilde{\omega}_{ab}&=& d\theta_{a-1} \left(\prod^{b-2}_{c=a}\sin{\theta_{c}} \right)\cos{\theta_{b-1}}\quad (a=1,\cdots, n-3),\quad (b=a+1,\cdots, n-2)\nonumber. 
\end{eqnarray}

Nontrivial components of the curvature tensor are
\begin{eqnarray}\label{a9}
R_{01}&=&-\frac{\eta}{R^{2}}e^{0}\wedge e^{1}, \nonumber\\
R_{0a}&=&-\frac{\eta}{R^{2}}e^{0}\wedge e^{a},\nonumber\\
R_{1a}&=& -\frac{1}{R^{2}}e^{1}\wedge e^{a}+ \frac{\cosh{\lambda}}{R} \left(\omega_{abc}- \frac{1}{2}C_{abc} \right) e^{b}\wedge e^{c}=-\frac{1}{R^{2}}e^{1}\wedge e^{a},\nonumber\\
R_{ab}&=& \frac{1}{R^{2}\sinh^{2}{\lambda}} \left(\tilde{R}_{a-1\,b-1\,c-1\,d-1}-\cosh^{2}{\lambda} \left(\delta_{ac}\delta_{bd}-\delta_{ad}\delta_{bc} \right)  \right), 
\end{eqnarray}
where $2\le a,\,b,\,c,\,d\le n-1$. Since both AdS$_{n}$ and H$^{n}$ are coset spaces, Ricci tensor is proportional to the metric;
\begin{equation}\label{a10}
R_{00}=-(n-1) \frac{\eta}{R^{2}},\quad R_{11}=-(n-1) \frac{1}{R^{2}},\quad R_{ab}=-(n-1) \frac{\delta_{ab}}{R^{2}}.
\end{equation}

\vskip 0.2cm
\noindent
{\bf C.4 Spin Connections of M$5$-bubble Background}\\
\noindent

Let us be specific to the extremal M$5$-bubble case and obtain the explicit form of the spin connection components. In order to compare the result with that of the extremal M$5$-branes (whose near-horizon geometry is AdS$_{6+1}\times$S$^{4}$), we give the expressions of both cases together at every step. 

\subsubsection*{Spin Connections for AdS$_{6+1}$ and H$^{7}$}

From the foregoing results (\ref{globalviel}), we can easily spell out  the explicit forms of the vielbeins for AdS$_{6+1}$ and H$_{7}$ involved with the extremal M$5$ branes/bubbles;
\begin{eqnarray}\label{a11}
e^{0}&=&2 Q^{ \frac{1}{3}}dt\cosh{\lambda}\,, \nonumber\\
e^{1}&=&2 Q^{ \frac{1}{3}}d\lambda,\\
e^{a}&=&2 Q^{ \frac{1}{3}}d\theta_{a-2}\sinh{\lambda}\left(\prod^{4}_{b=a-1} \sin{\theta_{b}} \right)\,, \quad (a=2,\cdots,6).\nonumber
\end{eqnarray}
 The nontrivial spin connections are
\begin{eqnarray}\label{a12}
\omega_{01}&=&\eta\, dt \,\sinh{\lambda}, \nonumber\\
\omega_{1a}&=&-d\theta_{a-2}\cosh{\lambda}\prod^{4}_{b=a-1}\sin{\theta_{b}},\quad(a=2,\cdots,\,6)\\
\omega_{ab}&=&d\theta_{a-2}\,\cos{\theta_{b-2}} \prod^{a-1}_{c=b-3}\sin{\theta_{c}}, \quad (a=2,\cdots,\,5\,;\,\, b=3,\cdots,\,6\,;\quad a+1\le b),\nonumber
\end{eqnarray}
It should be understood that $\eta=\eta_{00}=-1$ for AdS$_{6+1}$ while it is $1$ for H$_{7}$.  

\subsubsection*{Spin Connections for S$^{4}$ and dS$_{4}$}

We again use the results (\ref{left1form}) and (\ref{standardcoord}) but with different angular coordinates $(\phi(\equiv\chi_{0}),\,\chi_{1},\,\chi_{2},\,\chi_{3})$ in order to avoid any confusion with the ones used for AdS$_{6+1}$ and S$^{4}$. We use the conventional angular coordinates for S$^{4}$;
\begin{eqnarray}\label{a13}
e^{a+7}&=&Q^{ \frac{1}{3}}d\chi_{a}\prod^{3}_{b=a+1}\sin{\chi_{b}}, \quad (a=0,\cdots,\,3).
\end{eqnarray}

According to the Eq. (\ref{left1form}), the spin connection components read as 
\begin{eqnarray}\label{a14}
\omega^{a+7}{}_{b+7}=d\chi_{a}\cos{\chi_{b}}\prod^{b-1}_{c=a+1}\sin{\chi_{c}},\quad  (a=0,\,1,\,2\,;\,\, b=1,\,2,\,3\,;\quad a<b).
\end{eqnarray}
For de Sitter case in (\ref{left1form}), we use $\zeta=-1$. The same results can be obtained via Wick rotation from those of the sphere case. The vielbein $e^{\natural}$ goes to $i\bar{e}^{\natural}$ upon Wick rotation, $\chi_{3}=i\psi+\pi/2$ (where we used $\natural$ to denote the tenth component). This is because $\bar{\eta}_{\natural\,\natural}=\bar{\eta}^{\natural\,\natural}=\zeta=-1$ while Euclidean flat metric has been used for the sphere. The spin connection $\omega_{ab}=\omega_{abc}\,e^{c}$, being constructed from the frame structure constants via  $\omega_{abc}=\left(C_{abc}-C_{bac}-C_{cab} \right)/2$, takes extra $i$ on every occurrence of the subindex `$\natural$' upon Wick rotation. In all, the above equations are rephrased in de Sitter case as
\begin{equation}\label{a15}
\bar{e}^{a+7}=Q^{ \frac{1}{3}}d\chi_{a}\cosh{\psi}\prod^{2}_{c=a+1}\sin{\chi_{c}},\qquad \bar{e}^{\natural}=Q^{ \frac{1}{3}}d\psi.
\end{equation}
and the spin connection components become 
\begin{eqnarray}\label{a16}
\bar{\omega}^{a+7}{}_{b+7}&=&d\chi_{a}\cos{\chi_{b}}\prod^{b-1}_{c=a+1}\sin{\chi_{c}},\quad\\
\bar{\omega}^{a+7}{}_{\natural}&=&d\chi_{a}\sinh{\psi}\prod^{2}_{c=a+1}\sin{\chi_{c}},\quad (a=0,\,1,\,2\,;\quad b=1,\,2\,;\quad a<b).\nonumber
\end{eqnarray}

 The curvature tensor two forms are simply given by $R^{ab}=e^{a}\wedge e^{b}/Q^{ \frac{2}{3}}$ (and $\bar{R}^{ab}=\bar{e}^{a}\wedge \bar{e}^{b}/ Q^{ \frac{2}{3}}$ for de Sitter), thus with Ricci tensor components as $R_{ab}=3\delta_{ab}/Q^{ \frac{2}{3}}$ for a sphere and  $\bar{R}_{ab}=3\bar{\eta}_{ab}/Q^{ \frac{2}{3}}$ for de Sitter.
 
\subsubsection*{The explicit forms of the matrices $\Omega_{\mu}$}

The explicit forms of the matrices $\Omega_{\mu}$ defined in Eq. (\ref{kse}) are 
\begin{eqnarray}\label{a17}
\Omega_{t}&=& -\eta\left(\kappa\Gamma^{0789\natural}\cosh{\lambda}\, -\Gamma^{01}\sinh{\lambda}\,\right),\nonumber\\
\Omega_{\lambda}&=&-\kappa\Gamma^{1789\natural}, \nonumber\\
\Omega_{\theta_{a}}&=&-\left(\kappa\Gamma^{(a+2)789\natural}\sinh{\lambda}\,+\Gamma^{1(a+2)} \cosh{\lambda}\,\right)\left( \prod^{4}_{b=a+1}\sin{\theta_{b}}\right) +\Gamma^{(a+2)(a+3)}\cos{\theta_{a+1}}\nonumber\\
&&+\sum^{4}_{b=a+2}\Gamma^{(a+2)(b+2)}\cos{\theta_{b}}\prod^{b-1}_{c=a+1}\sin{\theta_{c}}, \quad (a=0,\cdots,3) \nonumber\\
\Omega_{\theta_{4}}&=& -\left(\kappa\Gamma^{6789\natural}\sinh{\lambda}\,+\Gamma^{16}\cosh{\lambda}\, \right),
\end{eqnarray}
for AdS$_{6+1}$/H$^{7}$ parts. 

For the sphere, the matrices are
\begin{eqnarray}\label{a18}
\Omega_{\phi} &=&\Gamma^{78}\cos{\chi_{1}}+\Gamma^{79}\sin{\chi_{1}}\cos{\chi_{2}}+\Gamma^{7\,\natural}\sin{\chi_{1}}\sin{\chi_{2}}\cos{\chi_{3}}+ \Gamma^{89\natural}\sin{\chi_{1}}\sin{\chi_{2}}\sin{\chi_{3}}, \nonumber\\
\Omega_{\chi_{1}}&=&\Gamma^{89}\cos{\chi_{2}}+\Gamma^{8\natural}\sin{\chi_{2}}\cos{\chi_{3}}-\Gamma^{79\natural}\sin{\chi_{2}}\sin{\chi_{3}},\nonumber\\
\Omega_{\chi_{2}}&=&\Gamma^{9\natural}\cos{\chi_{3}}+\Gamma^{78\natural}\sin{\chi_{3}},\nonumber\\
\Omega_{\chi_{3}}&=&-\Gamma^{789},
\end{eqnarray}
while for de Sitter part, they are
\begin{eqnarray}\label{a19}
\Omega_{\phi}&=&{\Gamma}^{78}\cos{\chi_{1}}+{\Gamma}^{79}\sin{\chi_{1}}\cos{\chi_{2}}+{\Gamma}^{7\,\natural}\sin{\chi_{1}}\sin{\chi_{2}}\sinh{\psi}+ i{\Gamma}^{89\natural}\sin{\chi_{1}}\sin{\chi_{2}}\cosh{\psi}, \nonumber\\
\Omega_{\chi_{1}}&=&{\Gamma}^{89}\cos{\chi_{2}}+{\Gamma}^{8\natural}\sin{\chi_{2}}\sinh{\psi}-i{\Gamma}^{79\natural}\sin{\chi_{2}}\cosh{\psi}, \nonumber\\
\Omega_{\chi_{2}}&=&{\Gamma}^{9\natural}\sinh{\psi}+i{\Gamma}^{78\natural}\cosh{\psi},\nonumber\\
\Omega_{\psi}&=&-i{\Gamma}^{789}.
\end{eqnarray}

\end{appendix}

%\bibliography{bubble}
%\bibliographystyle{abbrv}

\end{document}